# Reasoning about Interference Between Units

Jake Bowers [*]     Mark Fredrickson [†]     Costas Panagopoulos [‡]

July 13, 2012(Version 030f62f)

**Abstract**

If an experimental treatment is experienced by both treated and control group units, tests of hypotheses about causal effects may be difficult to conceptualize let alone execute. In this paper, we show how counterfactual causal models may be written and tested when theories suggest spillover or other network-based interference among experimental units. We show that the "no interference" assumption need not constrain scholars who have interesting questions about interference. We offer researchers the ability to model theories about how treatment given to some units may come to influence outcomes for other units. We further show how to test hypotheses about these causal effects, and we provide tools to enable researchers to assess the operating characteristics of their tests given their own models, designs, test statistics, and data. The conceptual and methodological framework we develop here is particularly applicable to social networks, but may be usefully deployed whenever a researcher wonders about interference between units. Interference between units need not be an untestable assumption; instead, interference is an opportunity to ask meaningful questions about theoretically interesting phenomena.

---

[*]Assistant Professor, Dept of Political Science, University of Illinois @ Urbana-Champaign (jwbowers@illinois.edu). *Acknowledgements:* Part of this work funded by NSF Grants SES-0753168 and SES-0753164 (joint with Ben Hansen, Dept of Statistics, University of Michigan). Thanks are due to the participants in the Student-Faculty Workshop in the Dept of Political Science at the University of Illinois at Urbana-Champaign, at the Experiments in Governance and Politics Workshop at MIT (EGAP-6), our panel at MPSA 2012, and SLAMM 2012. We especially appreciate the in-depth comments provided by Cyrus Samii, Cara Wong, Matthew Hayes, and the anonymous reviewers.

[†]Ph.D. Student in Political Science, MS in Statistics, University of Illinois @ Urbana-Champaign.
[‡]Assistant Professor, Dept of Political Science, Fordham University.

# 1 Introduction

In some experiments, intervention assigned to a treatment group is experienced by a control group. For example, in an election monitoring experiment, observers arriving at assigned-to-treatment villages find a peaceful election taking place in part because those aiming to unfairly influence the election saw the monitors arrive and took their intimidation elsewhere. If the thugs left a treatment village for a control village, what then is the causal effect of the election monitoring? A simple comparison of what we observe from treated and control villages would tell us about the impact of the treatment on *both* treated and control villages rather than how treatment changed the treated villages. What is more, if election monitoring is to be rolled out as a large-scale policy rather than as a field experiment, scholars need to assess models of the spillover of treatment from treated to control villages for the purposes of harnessing (or minimizing) that spillover. When policy makers ask whether road networks connecting villages would enable more movement of intimidation, or whether certain aspects of village social structure would make them easier or harder targets, researchers must be able to provide clear statements about the role of spillover in the proposed policy.

In this paper, we describe a way to formalize models that include explicit spillover between units. These models are theoretically driven. The method we propose provides wide latitude for researchers to transcribe the theoretical story of why subjects in a study react the way that they do, including reacting to the treatment of their neighbors. After writing down a theoretically driven model in a succinct mathematical form, we connect these models to the potential outcomes generated in the experiment. We show how to derive hypotheses from the models and how to test the hypothesis against experimental data. Throughout the paper, we use a simulated data set in which treatment spills over from treated to control units via a known network. Our simulated data set allows us to demonstrate that our method does not mislead researchers into rejecting true hypotheses, or failing to reject false hypotheses, too often. These results apply to hypotheses that include spillover and those that do not. Also, by varying the parameters of the simulation, we demonstrate techniques that researchers can use when designing studies and models to improve the efficiency of their designs.

By building on the testing tradition in statistics pioneered by Fisher (1935) and further developed most intensively by Rosenbaum (2002, 2010), we contribute to the methodological literature on the analysis of data from experiments. We show that one can model the flow of causal effects across networks, and that one can test the parameters of such models without requiring probability models of observed outcomes. We also remind political scientists that the "no interference" assumption so often invoked in discussions of causal inference is not a fundamental assumption, but an implication of the simple use of averages to conceptualize causal effects (Aronow, 2012; Rosenbaum, 2007).

By posing and testing hypotheses about interference, our effort is different from, yet complements, efforts that have mostly aimed at credible statistical inference built on an estimation framework that focuses on average treatment effects. There are many variants of this approach, yet they all involve a decomposition of average treatment effects into parts due to interference and parts not due to interference, often through clever research designs.[1] A very useful advance in this tradition, and a complement to our work, can be found in Aronow and Samii (2012a), where an approach to the estimation of average causal effects under general interference is developed. Our proposed framework, which we call "Fisherian" following Rosenbaum's terminology, builds a statistical methodology around testing, not estimation.[2] Although one rarely estimates without testing, a framework emphasizing testing has several advantages. Our approach does not require scholars to conceptualize causal effects in terms of averages, nor does it involve estimating effects. Our approach goes beyond tests of "no effects" to include theory driven statements about how direct and indirect effects might occur. Our approach is particularly useful when models of interference are complex, and when scientific interest does not focus on clarifying average differences but on other causal, counterfactual, quantities.

---

[1] For only a few examples of this approach, see McConnell, Sinclair and Green (2010); Sinclair (2011); Nickerson (2008, 2011); Hudgens and Halloran (2008); Sobel (2006); Tchetgen and VanderWeele (2010); VanderWeele (2008a,b, 2009, 2010); VanderWeele and Hernan (2011); Miguel and Kremer (2004); Chen, Humphreys and Modi (2010); Ichino and Schündeln (2011).

[2] Some of the point-estimation-based approaches cited above follow Neyman (1923 [1990]) in deriving the properties of estimators based on the randomized assignment of the experimental treatments and is known as "randomization inference." The testing approach we take in this paper follows Fisher (1935) and Rosenbaum (2002, 2010) and is also grounded in the randomized assignment of treatments, and is also known as "randomization inference." To avoid confusion in this paper, we talk about a "Fisherian" approach and try to avoid the term "randomization inference" unless we mean it to refer to both kinds of approaches. When causal models are simple and samples are large, the two kinds of randomization inference produce equivalent results and thus enable very convenient and fast computation (Samii and Aronow, 2011; Hansen and Bowers, 2008, 2009). For a general statement of this relationship between types of randomization-based inference, see Aronow

Two very different papers provide the proximate foundation for our work. Rosenbaum (2007) enables the production of confidence intervals for causal effects without assuming anything in particular about the form of interference between units. The key to his approach is the idea that the randomization distribution of certain distribution-free rank based test statistics can be calculated without knowing the distribution of outcomes (i.e. it can be calculated before the experiment has been run, when no unit at all has received treatment). Rosenbaum (2007) thus successfully enables randomization-justified confidence intervals about causal effects without requiring assumptions about interference. Our aim here, however, is more akin to Hong and Raudenbush (2006). They used a multilevel model to estimate the size of interference effects as they occurred between students nested within schools. We want to enable statistical inference about particular substantively relevant and theoretically motivated hypotheses about interference and causal effects simultaneously. Hong and Raudenbush (2006) also provide precedent for some of our work in collapsing aspects of the interference into a scalar valued function. Nothing about our approach requires us to collapse the possible avenues of interference in this way, but, in this, our first foray into asking questions about interference, such a simplification makes life much easier.

Our proposed framework can be applied to any experimental design, from simple to complex randomization schemes. Other, perhaps gold standard, approaches begin at the design stage and involve layered or multilevel randomization that directly assign units to "indirect" or spillover effects rather than direct effects (McConnell, Sinclair and Green, 2010; Hudgens and Halloran, 2008; Chen, Humphreys and Modi, 2010; Ichino and Schündeln, 2011). Our ideas here complement those designs by enabling scholars to specify more precisely mechanisms for both direct and indirect effects and

---

and Samii (2012*b*).

to assess the support in the data for such mechanisms. Our simulation examples demonstrate how design choices can influence experimental sensitivity, and we encourage scholars designing studies with a complex causal model of interference to replicate these simulations for their own context and model. We find that design choices regarding the model, the underlying distribution of data, the sample size, the network density, and even the percentage of treated units can influence the efficiency of an experimental design, sometimes in counter-intuitive ways.

Finally, as a paper written by social scientists rather than by statisticians, this contribution is not agnostic about the role of substantive theory in the enterprise of statistical inference about causal effects. That is, this paper considers interference between units as an implication of social and political processes to be reasoned about and tested. The conceptual framework and technology that allow us to engage so directly with interference build on Fisher's sharp null hypothesis and subsequent developments linking Fisher's original ideas with contemporary formal frameworks for conceptualizing causal effects and establishing statistical inferences (Fisher, 1935; Rosenbaum, 2002, 2010). We extend these developments to show that statistical inference about causally meaningful quantities is possible even when we hypothesize about interference directly. Social scientific theory can materially contribute to statistical inference via the specification of testable hypotheses, both those relating to interference and otherwise.

### 1.1 Roadmap

Throughout this paper we use a simulated experiment in which subjects are connected in a network. Treatment is randomly assigned to the subjects and an outcome is measured. In Section 2 we describe this simulated experiment and lay out our notation. In Sections 3 and 4, we demonstrate how to test hypotheses about treatment effects in this experiment, including hypotheses about explicit spillover between units. We focus on a small subsample of the entire experiment to make the process of writing microlevel models, and the implications for the potential outcomes of the subjects, more manageable. We then show how these hypotheses scale up to a more realistic sample size. In Section 5 we explore the operating characteristics of our method as we vary the parameters of the simulation. We show how changes in the information available influence a researcher's ability to reject false hypotheses (and also highlight how "information" includes but is not limited to "sample size"). We tackle the problem of comparing models and also show that researchers using this framework are unlikely to be misled into believing that spillover occurs when there is none. Section 6 considers issues of computational efficiency, the applicability of this method for observational data, and avenues for future research. Finally, in Appendix A, we show how to go from theory to code, both to enable the testing of hypotheses about interference, but also to assess the properties of any given model on any given dataset.

## 2 Setting

Imagine an experiment, perhaps similar to Panagopoulos (2006) or Ichino and Schündeln (2012), in which a randomized treatment has been allocated to units in a pre-existing network.[3] Figure 1 displays the network for such a simulated experiment graphically. This data set contains 256 subjects with 512 edges connecting the subjects. Treatment was randomly allocated to 128 of the 256 units.[4]

We will use this simulated network as a testbed for the rest of this paper. To introduce the foundations on which our method is built we now concentrate on a small subset of the data. Figure 2 highlights 7 subjects from the larger sample. This figure shows that a network can be represented both by a graph with nodes and edges as well as with a matrix. Throughout this paper we use an $n \times n$ adjacency matrix labeled $\mathbf{S}$ to record network relationships. An adjacency matrix contains a value of 1 at entry $i, j$ if there is a link between subject $i$ and subject $j$. In an undirected network, as shown here, a link from $i \rightarrow j$ implies a link from $j \rightarrow i$. For example, in Figure 2, the adjacency matrix shows $A \leftrightarrow D$ and $A \leftrightarrow g$ but no link between $A$ and $b$. We focus on models for undirected

---

[3]In Panagopoulos (2006)'s study of newspaper advertisements applied to US cities, it was reasonable to consider the possibility that citizens in nearby control cities read the same newspapers as citizens in treated cities. Ichino and Schündeln (2012)'s study of election observers was designed to detect positive spillovers, such that election observers may not be needed in every village in order to ensure clean elections.

[4]This network is like a road network in that it is relatively sparse: 50% of the nodes have between 2 and 5 edges directly connecting them with other nodes. Notice that the network as drawn may be fixed — like a road network — or may represent hypotheses (such that unit $A$ and $b$ could, in principle, interfere, but we have chosen not to model such a connection).

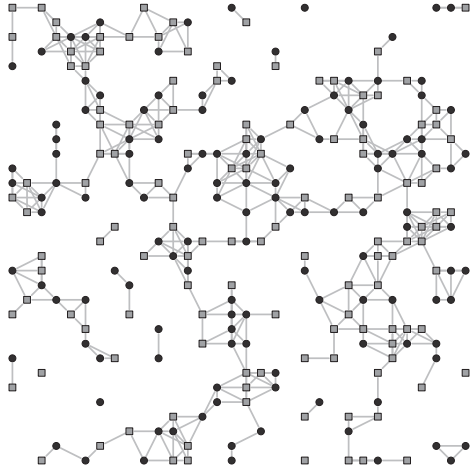

Figure 1: A simulated data set with 256 units and 512 connections. The 256/2 = 128 treated units are shown as filled circles and an equal number of control units are shown as as gray squares.

networks in this paper although the method easily extends to directed networks.

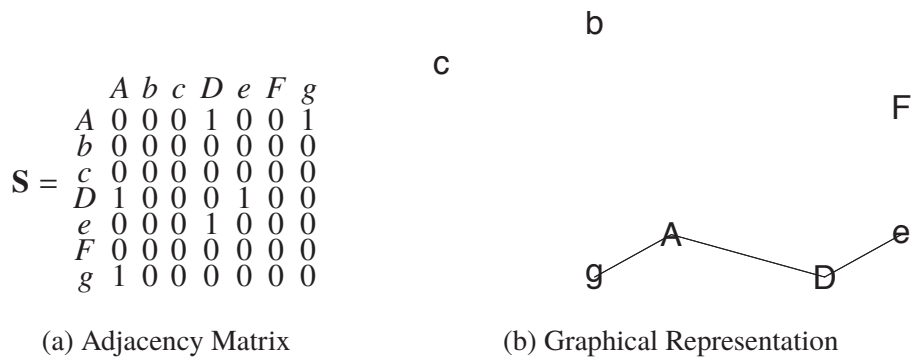

(a) Adjacency Matrix

(b) Graphical Representation

Figure 2: Graphical and adjacency matrix representations of connections for a subgroup of the large, dense network displayed in Figure 7. Capital letters indicate treated units, lower case letters control units.

Write $Z_i = 1$ to indicate that subject $i$ is assigned to the treatment condition and $Z_i = 0$ to indicate subject $i$ is assigned to the control condition and collect those indicators of treatment into a vector, $\mathbf{Z}_{n \times 1}$. There are $2^n$ possible unique vectors $\mathbf{Z}$. In our experiment, $\mathbf{Z}$ is generated by assigning half of the 256 subjects to each of the treatment and control conditions with equal probability. Let $\Omega$ be the sample space of treatment assignments. There are $\binom{256}{128} = |\Omega| = 5.77 \times 10^{75}$ ways that treatment could be allocated to this subject pool. In our small subset, with only 7 units, there are $\binom{7}{3} = 35$ ways that $\mathbf{Z}$ could be drawn from $\Omega$.[5]

---

[5]For the purpose of simplicity, we restrict our attention to binary treatments assigned to a fixed percentage of the sample (e.g. precisely 50% of the sample is treated in all $\mathbf{Z}$). Our method

We formalize the idea that the treatment assigned to the set of units caused some change for unit $i$ using the counterfactual notion of causal effects: a particular set of treatments applied to this network causes an effect if unit $i$ would have acted differently with this treatment assignment vector than it would have under a different set of assignments. We write $y_{i,\mathbf{z}}$ to represent the "potential outcome" that would be observed for unit $i$ if $\mathbf{Z} = \mathbf{z}$, and we write $\underset{n\times 1}{\mathbf{y_z}}$ to represent the potential outcomes of all units to treatment $\mathbf{Z} = \mathbf{z}$. We say treatment caused an outcome if $\mathbf{y_z} \neq \mathbf{y'_z}$, where $\mathbf{z} \neq \mathbf{z'}$.[6] In the potential outcomes framework there are $2^n$ possible outcomes for each unit, each corresponding to a unique value of $\mathbf{z}$. As we only allocate treatment in one arrangement to our subjects, we therefore only observe a single potential outcome for each unit in the sample. In the case of our small network shown above, the observed treatment allocation was $\mathbf{z} = \{1, 0, 0, 1, 0, 1, 0\}$. We therefore write the observed outcome as $\mathbf{y}_{1001010}$, the potential outcome for our observed treatment assignment.[7]

Two units can be said to *interfere* with each other when the potential outcomes of one unit depend on treatment assigned to another unit. Equivalently, two units interfere when the potential outcomes of one unit depend on the potential outcomes of another unit (since potential outcomes are defined in reference to treatment assignment). We write out the entire vector $\mathbf{z}$ in describing potential outcomes to highlight the possibility that the outcome for subject $i$ ($y_{i,\mathbf{z}}$) may depend on the treatment assignment of some other set of subjects. Readers may be accustomed to thinking about the causal effect of an experiment in terms of a comparison of the potential outcomes we would see for subject $i$ if that subject were treated, $y_{i,z_i=1} \equiv y_{i,1}$, and the outcome we would see if

---

applies equally well to multiple experimental levels and alternative randomization mechanisms (e.g., stratified randomization, independent coin flips, hierarchical and clustered randomizations).

[6] For an introduction to the idea of using of potential outcomes to formalize counterfactual notions of causal relations in political science see Sekhon (2008); Brady (2008). For a more general overview see Rubin (2005).

[7] It is more common to write $Y_i$ for an observed outcome that depends on the random assignment of treatment and $y_{i,Z_i}$ for the fixed potential outcome to treatment $Z_i$. Yet, with complex interference we have discovered that it is simpler to write $y_{i,Z_i=z_i}$ or "the potential outcome given the observed treatment assignment vector" as "the observed outcome" rather than to add an equation for $Y_i$ with possibly $2^n$ terms in need of substitution and solution.

treatment were withheld, $y_{i,z_i=0} \equiv y_{i,0}$. This notation implies no interference between units. To write $y_{i,z_i=1}$ or $y_{i,z_i=0}$ is to say that $y_{i,z_i,\mathbf{z}_{-i}} = y_{i,z_i,\mathbf{z}'_{-i}}$ for all $\mathbf{z} \neq \mathbf{z}'$ (where $\mathbf{z}_{-i}$ refers to the treatment vector excluding the entry for unit $i$). This manner of writing potential outcomes encodes a decision to focus attention only on unit $i$ and to exclude from consideration the treatment assignment status of other units. Much of the literature calls this notational choice a manifestation of an assumption of no interference. As we show in the following sections, an assumption of no interference is not a necessity of the potential outcomes framework, but an implication derived from modeling choices.[8]

## 3 Method: Hypotheses and Models

We define a *causal model* to be a function $\mathcal{H}(y_{i,\mathbf{z}}, \mathbf{w}, \theta) = y_{i,\mathbf{w}}$ that transforms a potential outcome for one treatment vector $\mathbf{z}$, $y_{i,\mathbf{z}}$, to the potential outcome for another treatment vector $\mathbf{w}$. In vector notation, we might replace $y_{i,\mathbf{z}}$ with the vector $\mathbf{y}_\mathbf{z}$ to indicate $\mathcal{H}$ is applied to the entire sample with the same $\mathbf{z}$ and $\mathbf{w}$ arguments.[9] The parameter $\theta$ defines the *causal effect* of the model and serves to generate specific hypotheses, which we demonstrate in more detail in § 4.

To make these definitions more concrete, consider the simplest model: that the treatment assignment had no causal effect on any unit, often called the "sharp null hypothesis of no effects." This model states that any treatment assignment would not change the outcome of any subject in the experiment:

---

[8]Many expositions of the potential outcomes formalization of causal inference make the "no interference" assumption in the context of a broader umbrella called the Stable Unit Treatment Value Assumption (or SUTVA) (See, Brady, 2008, for a discussion aimed at political scientists). Whereas Cox (1958, § 2.4) discusses the need to assume no interference in order to reason simply about average treatment effects, Rubin (1980, 1986) amplifies this requirement to also include the idea that treatment assigned to one unit is the same as the treatment assigned to another unit (i.e. there are no "varieties" or "types" of treatment that are not recorded in our treatment assignment vector).

[9]When considering models that include an assumption of no interference, we will write $\mathcal{H}(y_{i,z_i}, w_i) = y_{i,w_i}$ to indicate that the entries $z_j \in \mathbf{z}$ and $w_j \in \mathbf{w}$ for $j \neq i$ do not change the potential outcomes for subject $i$.

$$\mathcal{H}(\mathbf{y_z}, \mathbf{w}) = \mathbf{y_z} \tag{1}$$

By definition $\mathcal{H}(\mathbf{y_z}, \mathbf{w}) = \mathbf{y_w}$, therefore this model states that $\mathbf{y_z} = \mathbf{y_w}$. As the sharp null does not make use of parameters, we omit $\theta$ when discussing the sharp null of no effects.

Let $\mathbf{z} = \mathbf{0} = \{z_1 = 0, \ldots, z_n = 0\}$, the treatment assignment vector in which all units receive the control condition (i.e., $\mathbf{z}$ is all zeros). The potential outcome to this condition is written $\mathbf{y_0}$. We call this baseline condition the "uniformity trial" following Rosenbaum (2007).[10] When the control condition involves no action by the researcher, we can think of the uniformity trial as the world we would have observed if no experiment had been carried out at all. In many experiments, the treatment condition is compared to a standard procedure. For example, drug trials compare the efficacy of new drugs to the currently available prescription. In these cases, the uniformity trial is the world in which all subjects received the established drug. Using the uniformity trial, we see that we could write Equation 1 as $\mathcal{H}(\mathbf{y_0}, \mathbf{w}) = \mathbf{y_0}$. In other words, for any treatment assignment $\mathbf{w}$, the potential outcome $\mathbf{y_w} = \mathbf{y_0}$.

One often encounters this model labeled as a hypothesis: $H_0 : y_{i,\mathbf{z}} = y_{i,\mathbf{0}}$. Yet, when we consider complex models of interference between units it will be useful for us to think of hypotheses stated in terms of potential outcomes as generated and made meaningful by models. This model says that the way that unit $i$ reacted to the treatments given to the network would be the same as the way it would react to the situation where no treatment is given to any member of the network—where the experiment is not even fielded. In the Fisherian framework, we can test models in which potential outcomes are defined with respect to a causal model $\mathcal{H}$, even if we imagine interference among potential outcomes. (Rosenbaum, 2007; Aronow, 2012). Here we briefly describe how to test the sharp null of no effects even considering interference between units using our simulated data.[11]

---

[10]Rosenbaum (2007) adopts the term from the name of a method used to calibrate variance calculations in agricultural experiments by assigning treatment but not applying it.

[11]We present here a very condensed summary of the Fisherian framework for statistical inference. For more detail see Keele, McConnaughy and White (2012), Rosenbaum (2010, Chap 2), and Rosenbaum (2002, Chap 2).

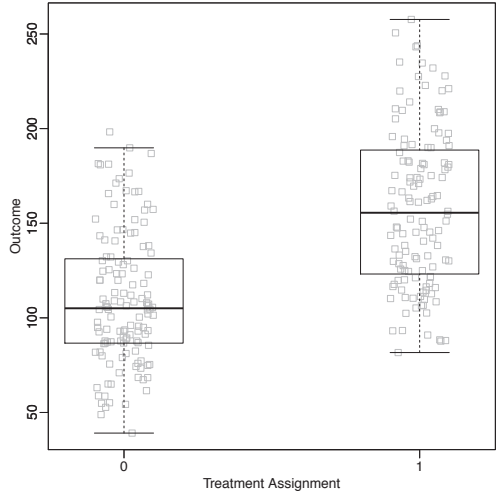

Figure 3: The observed outcome, $\mathbf{y_z}$, for the simulated data set with 256 units and 512 connections, broken out for treated and control units.

Figure 3 shows the measured outcome distribution for the treated and control groups in the 256 subject experiment, from Figure 1. What ought we to observe if our model, $\mathcal{H}(\mathbf{y_0}, \mathbf{w}) = \mathbf{y_0}$, held? If treatment had no effect, we would expect the treated and control groups to be random samples from a common distribution. If the sharp null hypothesis were true, the two box plots in Figure 3 should look the same except for a little noise arising from the sampling procedure. Conversely, if the model were wrong, and this experimental manipulation had a causal effect, we would expect these two distributions to differ in a systematic way.

To score the dissimilarity of treated and control distributions we employ a test statistic, which we notate $\mathcal{T}(\mathbf{y_0}, \mathbf{z})$. The value $\mathcal{T}$ should small when treated and control distributions are similar and large when they are dissimilar. Although it is common to use differences in means or differences in sums of ranks to compare distributions, in this paper we rely on the Kolmogorov-Smirnov (KS) test statistic. As will become clear later when models are more complex, we want a test statistic that is sensitive to differences in distribution that include diffrences in center as well as differences in spread, skew, other aspects of distributions.[12] Figure 3 displays both a difference in center and an increase in spread for the treated group relative to the control group.

---

[12]The KS test statistic is the maximum difference between the empirical cumulative distribution function (ECDF) of the treated units $F_1$ and the ECDF of the control units $F_0$: $\mathcal{T}(\mathbf{y_0}, \mathbf{z}) = \max_{i=1,\ldots,n} [F_1(y_{i,\mathbf{0}}) - F_0(y_{i,\mathbf{0}})]$, where $F(x) = (1/n) \sum_{i=1}^{n} I(x_i \leq x)$ tells us the proportion of the distribution of $x$ at or below $x_i$ (Hollander, 1999, §5.4).

Consider a replication of the experiment, where everything is the same except that we choose another $\mathbf{z}'$ from $\Omega$, where $\mathbf{z} \neq \mathbf{z}'$. If the sharp null of no effects model were true, we would observe the same outcome in each trial $\mathbf{y_z} = \mathbf{y_{z'}} = \mathbf{y_0}$. However, the value of the test statistics $\mathcal{T}(\mathbf{y_0}, \mathbf{z})$ and $\mathcal{T}(\mathbf{y_0}, \mathbf{z}')$ might differ because of chance variation induced by the values of $\mathbf{z}$ and $\mathbf{z}'$. Fisher (1935, Chap 2) showed how to generate this distribution by enumerating the possible ways for the experiment to occur: By computing the value of $\mathcal{T}(\mathbf{y_0}, \mathbf{Z})$ for every possible treatment assignment $\mathbf{Z} \in \Omega$, we can generate the randomization distribution of the test statistic under the null hypothesis of effects. The $p$-value of the hypothesis is defined as $\Pr(\mathcal{T}(\mathbf{y_0}, \mathbf{Z}) > \mathcal{T}(\mathbf{y_0}, \mathbf{z}))$, where $\mathbf{z}$ is the observed treatment assignment. Recall, that according to our model, $\mathcal{H}(\mathbf{y_z}, \mathbf{0}) = \mathbf{y_z} \equiv \mathbf{y_0}$ for observed data $\mathbf{y_z}$ and treatment assignment $\mathbf{z}$. The causal model $\mathcal{H}$ links the observed data with the uniformity trial, and the research design and scores for $\mathcal{T}$ then imply a clear distribution for $\mathcal{T}$. Large-sample normal approximations also exist for many test statistics including the KS test statistic, and Figure 4 uses the large-sample approximation to show the distribution under the model of no effects using the data shown in Figure 3. For the example data, the sharp null of no effects has a $p$-value of $6.83 \times 10^{-14}$, suggesting that there is very good evidence against the hypothesis generated by the model of no effects.

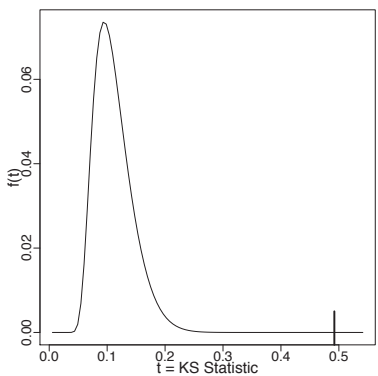

Figure 4: The large-sample distribution of the Kolmogorov-Smirnov test statistic for the example 256 subject experiment under the model $\mathcal{H}(\mathbf{y_z}, \mathbf{0}) = \mathbf{y_z} \equiv \mathbf{y_0}$. The vertical black line represents the value of the test statistic for the observed data $\mathbf{y_z}$. The $p$-value ($6.83 \times 10^{-14}$) is computed by taking the area under curve to the right of the vertical line representing $\mathcal{T}(\mathbf{y_z}, \mathbf{z})$.

In the preceeding discussion of the sharp null, we have made no assumption about whether interference exists or not. On one hand, all potential outcomes could be the same because the treatment has no direct effect and there are no spillover effects. The model of no effects may also be true because direct and spillover effects precisely cancel. These two scenarios are observationally equivalent: no matter how we assign treatment, we would measure the same outcome. But the two processes generating these outcomes are quite different. In the next section, we address how to go beyond the sharp null of no effects and demonstrate how the Fisherian testing framework can help researchers model these underlying processes, including processes that lead to spillover between units.

## 4 Models of interference

Asking questions about "no effects" is only the beginning. In this section we describe how to write down and test causal models of interference between units. We focus attention on one particular model to formalize an intuitive story of how treatment "spills over" from treated to control subjects. We also demonstrate the use of Hong and Raudenbush's (2006) simplifying idea of reducing the network effects to a scalar quantity. We begin by outlining several points as plausible grounds for a theoretically driven model. We then describe the theoretically derived aspects of our model and show how we capture them in a mathematical causal model of the form $\mathcal{H}(\mathbf{y_z}, \mathbf{z}, \mathbf{w}, \theta)$.

The core elements of our illustrative substantive theory are:

1. Interference between units is possible, but interference is limited to the network shown in Figure 1. We represent the network as the adjacency matrix $\mathbf{S}$, where a link between units is represented by the value 1, and zero otherwise.

2. The spillover effect depends only on the number of neighbors treated. We compute the number of treated neighbors for each unit using the expression $\mathbf{z}^T\mathbf{S}$. This expression collapses the network's influence into a scalar summary following Hong and Raudenbush (2006).

3. The treatment has a direct effect on the treated units. The direct effect is greater for units that would have had a larger outcome if treatment hadn't been applied to any unit (i.e. a larger or higher outcome in the uniformity trial). This concept implies a multiplicative effect, which we parameterize as $\beta$.

4. This direct effect is always greater in magnitude than the spillover effect. To limit the magnitude of the spillover, we employ a non-linear growth curve expression:

$$\beta + (1 - \beta) \exp\left(-\tau^2 \mathbf{z}^T \mathbf{S}\right). \tag{2}$$

The value of this function will always remain between 1 and $\beta$, with a rate of growth controlled by $\tau$, the second parameter in our model. Figure 5 shows how this expression behaves for different values of $\tau$ as the number of treated neighbors increases.

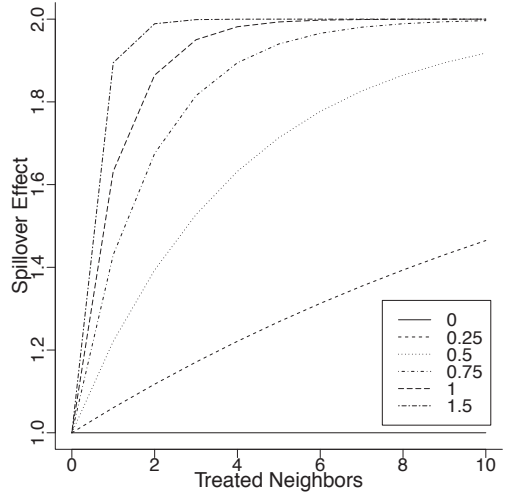

Figure 5: Growth curve of spillover effects for the expression $\beta + (1 - \beta) \exp\left(-\tau^2 \mathbf{z}^T \mathbf{S}\right)$ as the number of treated neighbors, $\mathbf{z}^T \mathbf{S}$, increases for $\beta = 2$ and a selection of $\tau$ values.

5. Spillover only flows from treated units to control units, and not from any other combination. Thus treated units will not experience a spillover effect from treated neighbors. When units are treated, they will only experience a multiplicative effect $\beta$, regardless of their locations in the network. We can extend the growth curve expression slightly to filter out the spillover effect for units where $z_i = 1$:

$$\beta + (1 - z_i)(1 - \beta) \exp\left(-\tau^2 \mathbf{z}^T \mathbf{S}\right) \tag{3}$$

Since $\mathcal{H}$ is a function that transforms one potential outcome into another, we begin by considering the transformation from the uniformity trial $\mathbf{y_0}$ to the outcome we would observe with treatment $\mathbf{z}$: $\mathcal{H}(\mathbf{y_0}, \mathbf{z}, \theta) = \mathbf{y_z}$.

$$\mathcal{H}(\mathbf{y_0}, \mathbf{z}, \beta, \tau) = \left[\beta + (1 - z_i)(1 - \beta) \exp\left(-\tau^2 \mathbf{z}^T \mathbf{S}\right)\right] \mathbf{y_0} \tag{4}$$

Equation 4 states that when we apply treatment $\mathbf{z}$ to the uniformity trial, treated units see their outcomes multiplied by $\beta$, while control units experience a spillover effect that increases up to $\beta$ as the the number of treated neighbors increase. By definition, $\mathcal{H}(\mathbf{y_0}, \mathbf{z}, \beta, \tau) = \mathbf{y_z}$. By replacing the left hand side with $\mathbf{y_z}$ and solving for $\mathbf{y_0}$, we can write $\mathcal{H}$ to transform the observed data into the uniformity trial:

$$\mathcal{H}(\mathbf{y_z}, \mathbf{0}, \beta, \tau) = \left[\beta + (1 - z_i)(1 - \beta) \exp\left(-\tau^2 \mathbf{z}^T \mathbf{S}\right)\right]^{-1} \mathbf{y_z} \equiv \mathbf{y_0} \quad (5)$$

Combining these two equations for specific transformations gives us the general form of $\mathcal{H}$ for our model as:

$$\mathcal{H}(\mathbf{y_z}, \mathbf{w}, \beta, \tau) = \frac{\beta + (1 - w_i)(1 - \beta) \exp(-\tau^2 \mathbf{w}^T \mathbf{S})}{\beta + (1 - z_i)(1 - \beta) \exp(-\tau^2 \mathbf{z}^T \mathbf{S})} \mathbf{y_z} \quad (6)$$

Notice that this model contains other, simpler models nested within it. When $\tau = 0$, this model reduces to a model that implies no spillover effects, only a multiplicative direct effect. When $\beta = 1$, this model reduces to the sharp null of no effects, regardless of the value of $\tau$.[13]

Models with an interference parameter ($\tau$) and a causal effect parameter ($\beta$) seem to arise naturally when we consider interference between units. Statistical inference in the presence of parameters like $\tau$ depends on one's perspective on $\tau$. If $\tau$ is a fixed feature of the design, inference may proceed by setting such parameters at fixed values, or one may consider $\tau$ as a kind of tuning parameter, and values for it could be chosen using a power analysis or cross-validation. If $\tau$ is not fixed but is considered a nuisance parameter, then one can produce confidence intervals either by (1) assessing a given hypothesis about $\beta$ over the range of $\tau$, keeping the hypothesis about $\beta$ with the largest $p$-value (Barnard, 1947; Silvapulle, 1996) or (2) producing a confidence interval for $\tau$ and adjusting the largest $p$-value from a set of tests about a given $\beta_0$ over the range of $\tau$ in the confidence interval (Berger and Boos, 1994; Nolen and Hudgens, 2010). Nolen and Hudgens (2010) show that either solution will maintain the correct coverage of the resulting confidence intervals about treatment effects, although using the largest $p$-value is apt to make those confidence intervals overly conservative. In this paper we take a different approach: Parameters like $\tau$ need not be a nuisance.

---

[13]Readers may be more familiar with causal models of effects in the form of an identity equation, rather than a function that transforms between potential outcomes. For example, an additive model without spillover is often written as $y_{i,1} = y_{i,0} + \tau$ or $H_0 : y_{i,1} = y_{i,0} + \tau$. We can translate this notation to our own by replacing the explicit potential outcomes with $z_i$: $y_{i,z_i} = y_{i,0} + z_i \tau$. This second identity yields the proper outcome when $z_i = 1$ and $z_i = 0$. We can replace the left hand side with the model description: $\mathcal{H}(y_{i,z_i}, 0, \tau) = y_{i,0} + z_i \tau$. As with our spillover example, we can solve the previous identity for $y_{i,0}$ to get $\mathcal{H}(y_{i,0}, z_i, \tau)$. Combining the two specific forms of $\mathcal{H}$ into a general statement yields: $\mathcal{H}(y_{i,z_i}, w_i, \tau) = y_{i,z_i} + (w_i - z_i)\tau$. When models only consider two potential outcomes per unit, $y_{i,1}$ and $y_{i,0}$, the identity notation is certainly very convenient. As shown here, our notation encompasses the identity. We believe our notation has the added advantage of succinctly covering models of spillover, where the identity style notation quickly becomes cumbersome.

In this case, we wrote our model so that $\tau$ represents the rate of spillover. In fact, we can easily assess hypotheses about $(\beta, \tau)$ pairs using the same technique introduced in the previous section. We showed how the sharp null hypothesis of no effects can be tested by computing the distribution of $\mathcal{T}(\mathbf{y_0}, \mathbf{Z})$, the test statistic applied to the uniformity trial $\mathbf{y_0}$ suggested by the null hypothesis with respect to the randomly allocated treatment assignment $\mathbf{Z}$. In Equation 5, we have shown how to write a theoretically driven model of spillover that transforms the potential outcomes observed into the uniformity trial implied by the model. To test the hypothesis $H_0: \beta = \beta_0, \tau = \tau_0$, we substitute the hypothesized values into the adjustment equation to get $\mathbf{y_0} = \left[\beta_0 + (1 - z_i)(1 - \beta_0) \exp\left(-\tau_0^2 \mathbf{z}^T \mathbf{S}\right)\right]^{-1} \mathbf{y_z}$, with observed data $\mathbf{y_z}$. As before, we again compute the $p$-value of the hypothesis as $\Pr(\mathcal{T}(\mathbf{y_0}, \mathbf{z}) > \mathcal{T}(\mathbf{y_0}, \mathbf{Z}))$. If $\mathcal{T}(\mathbf{y_0}, \mathbf{z})$ is small it suggests that after adjustment the treated and control groups appear to come from the same distribution. If the $\mathcal{T}(\mathbf{y_0}, \mathbf{z})$ is large, it suggests that the model does not do a good job of describing the effect of treatment. The $p$-value quantifies this discrepancy with respect to the values of $\mathcal{T}$ we would expect to see if the null hypothesis were true.

Figure 6 shows a plot of the $p$-values for a series of $(\beta, \tau)$ hypotheses tested using the KS test statistic as applied to $\mathcal{H}(\mathbf{y_z}, \mathbf{0}, \beta, \tau)$ for a range of $\beta, \tau$ pairs, where $\mathbf{y_z}$ and $\mathbf{z}$ are the observed outcome and treatment assignment shown in Figure 3. The figure shows that as we begin to entertain hypotheses about some positive amount of spill-over, the confidence interval for $\beta$ expands. This is sensible: if, when treatment is assigned to one unit, most of that treatment is also experienced by another unit, then we have less information available about the treatment effect than we would have if the two units had been independent. Consider the extreme case in which treatment assigned to one unit is fully experienced by the relevant control unit — then we would not have enough information to calculate a treatment effect at all and any reasonable procedure should produce an infinite interval. The hypotheses about which we have least doubt (or highest $p$-values) cluster around the value which produced the data ($\tau = 0.5, \beta = 2$).

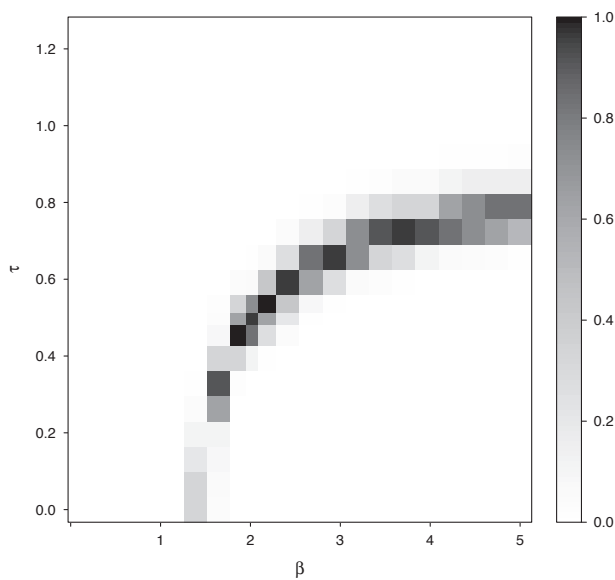

Figure 6: $p$-values for the model in Equation 6 for a series of joint hypotheses over $\beta$ and $\tau$. The true values are $\tau = 0.5$ and $\beta = 2$.

This simple illustration shows that models of interference may be conceptualized and tested in Fisher's framework. It is not an argument in favor of a particular model. The main point is that one can reason directly about interference and such reasoning, if formalized, can produce hypotheses about both causal effects and structural features of the effects. The data can provide evidence against such hypotheses. Multi-dimensional hypotheses can be tested to produce substantively interesting and useful $p$-value regions. The region tells the analyst both about what kinds of values are implausible under a theoretically informed model and also about the amount of information available to make such plausibility assessments.

### 4.1 A Fisherian Inference Algorithm

The Fisherian framework can be applied to models that include effects due to interference and models that assume no interference. In the following algorithm, we detail the procedure for testing a model $\mathcal{H}$ against observed data, with special attention to interference.

1. Write down a causal counter-factual model that specifies relations among potential outcomes, $\mathcal{H}(y_{i,\mathbf{u}}, \mathbf{w}, \theta)$ (where $\mathbf{u}$ and $\mathbf{w}$ are arbitrary treatment assignments). Include the treatment assignments of $u_j$ and $w_j$ for $j \neq i$ if spillover effects are theoretically motivated.

2. Map the causal model to the observed outcomes ($\mathbf{y_z}$) and treatment ($\mathbf{z}$) to the uniformity trial: $\mathcal{H}(\mathbf{y_z}, \mathbf{0}, \theta) = \mathbf{y_0}$ for a hypothesized value of $\theta$.

3. Select a test statistic $\mathcal{T}$ that will be small when the treated and control distributions in the adjusted data from step (2) are similar, and large when the distributions diverge.

4. Generate the distribution of $\mathcal{T}(\mathbf{y_0}, \mathbf{Z})$ under the hypothesis created in (2). The exact distribution arises from computing $t_k = \mathcal{T}(\mathbf{y_0}, \mathbf{Z}_k)$ for each $\mathbf{Z}_k \in \Omega$. Alternatively, sample from that distribution or approximate it using limit theorems.

5. The $p$-value for the specific hypothesis generated from the model is
$$\frac{\sum_{k=1}^{|\Omega|} \mathrm{I}(\mathcal{T}(\mathbf{y_0}, \mathbf{z}) > t_k)}{|\Omega|} \tag{7}$$

For a model that includes parameters, this process can be repeated for each unique parameter combination. One can summarize these $p$-values with intervals or regions for which the hypotheses are not rejected at a given $\alpha$-level.

In this section we demonstrated writing down a concrete model of effects with explicit spillover effects. Our model used parameters $\beta$ and $\tau$, and we showed that each pair of parameters formed a testable hypothesis. While we have demonstrated that testing hypotheses generated by such models is possible, readers may be left wondering about how this method performs. The key question, from the perspective of evaluating a statistical procedure is about how our framework's operating characteristics change as information changes: as sample size increases, does it more easily reject incorrect hypotheses? Do changes in network density, treatment assignment proportion, or test statistic choice also matter for the power and/or error rate of the procedure? In the next section, we engage with these questions to provide some guidelines for the use of this approach in analysis as well as in design.

## 5  Operating Characteristics of Tests for Models of Interference

Our approach allows models of the substance of the causal process, including interference, to have direct implications for data. Yet, data also have implications for tests. A well operating test should, at minimum, have two characteristics: (1) When the test faces a true hypothesis, it should rarely produce low *p*-values. That is, a test of a true hypothesis should encourage us to reject the hypothesis rarely, and, moreover, have a controlled error rate in this regard. This characteristic is sometimes called "correct coverage" or "unbiasedness of tests": "An ($\alpha$)level test is unbiased against a set of alternative hypotheses if the power of the test against these alternatives is at least ($\alpha$)" (Rosenbaum, 2010, Glossary). (2) When a test faces a false hypothesis—a hypothesis which is incongruent with the processes generating the data—the test should produce small *p*-values. That is, a test of a false hypothesis should encourage us to reject hypotheses far from the data often. This characteristic is called "power". If a test has sufficient power, it is a "consistent test": "A consistent test is one that gets it right if the sample size is large enough. A test of a null hypothesis ($H_0$) is consistent against an alternative hypothesis ($H_A$) if the power of the test tends to one as the sample size increases when ($H_A$) is true" (Rosenbaum, 2010, Glossary). It is common practice to provide analytic proofs of the asymptotic properties of new procedures. Yet, a researcher with her own model, design, and data never knows whether such a proof applies (and how) in her own case. To focus squarely on aid to applied researchers, we here present a series of simulation studies to assess the operating characteristics of the algorithm described in § 4.1 using the model specified in equation 6. In addition, statistical inference for experiments on networks raises questions about what "asymptotic" might mean: information about hypotheses might rise and fall with sample size, but also with density of network connections, and proportion assigned to treatment (if not also with test statistic choice and the extent to which the network drives the baseline outcomes). Our simulation framework enables us to explore all of the different ways in which the information available to a test may vary.

In what follows, we compare the size of a test (the probability that it rejects a correct hypothesis) with its level (the pre-specified probability of rejecting a correct hypothesis). Rosenbaum (2010, Glossary) calls the level of a test the "promise" of a given error rate. So, this assessment can be thought of as asking whether a given hypothesis testing procedure fulfills (size) its promises (level). In our simulations, we demonstrate that our method keeps its promises. We also assess the test's power to reject false hypotheses. As we test hypotheses that diverge from the truth, we should prefer models that allow us to reject these hypotheses at a high rate. In our simulations, we observe that test power does not always behave as expected. Increasing sample size leads to increased power, as we might expect, but due to the complex interaction of the model and the fixed network, there can be non-monotonic relationships with network density and the percentage of units that are treated.[14]

In the previous section, we used a theoretically derived model to show that reasoning about interference was possible and encouraged researchers to write down their own models of interference. In this section, we demonstrate simulation techniques for our model and again encourage researchers to follow our lead with respect to assessing the characteristics their own models and experimental designs. To further aid applied researchers, all simulations in this section and all hypothesis tests in Sections 3 and 4 are based on the development version of our freely available, open-source software package, RItools, available at https://github.com/markmfredrickson/RItools. Appendix A contains annotated code fragments used in this paper, and the source code to the entire document is available in the online supplemental materials at [URL].

### 5.1 Simulation setup

Our canonical experiment involves 256 subjects with 512 edges connecting them using the data already discussed in Section 2. To maintain continuity across simulations that vary sample size and network density, we employ the following rules when generating the simulated data. First, there is a pre-specified order in which units are added to the sample. This order is randomly generated, but is the same for all simulations. Therefore, all simulations with the same sample size use the same units.

---

[14]We thank an anonymous reviewer for bringing this second relationship to our attention.

Second, units have a pre-specified location on the grid, and edges are placed between closest pairs first. Figure 7 shows three networks that vary on sample size and the total number of edges. These three plots also demonstrate example treatment assignments (circles indicate the control condition, squares the treatment condition). The rightmost panel is the network used in the previous examples.

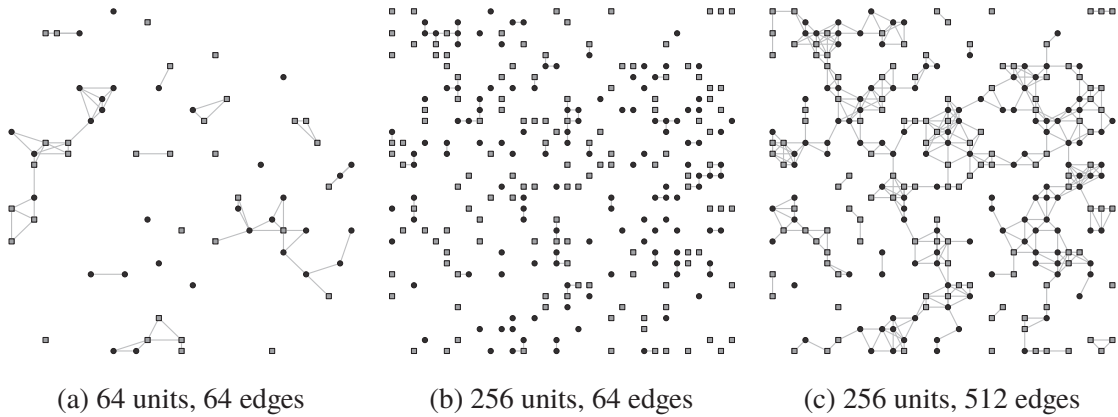

(a) 64 units, 64 edges     (b) 256 units, 64 edges     (c) 256 units, 512 edges

Figure 7: Example undirected, random, social networks assigned to treatment (circle) or control (square).

Unless otherwise noted, we use the model shown in equation 6 to generate the observed data from a fixed uniformity trial data set. We fix the values $\tau = 0.5$ and $\beta = 2$.

Our simulation study uses 1000 repetitions of the following algorithm:

1. Generate the uniformity trial data, the world in which the experiment was never fielded.

2. Draw a vector of treatment assignments from the set of possible assignments consistent with the design.

3. Generate a set of observed outcomes from the fixed uniformity trial following the treatment assignment and the spillover model in Equation 6.

4. Test hypotheses using the algorithm listed in Section 4.1. A hypothesis is any model and parameter set. For example, the true functional form with the true parameters $\tau = 0.5$ and $\beta = 2$ is one hypothesis, in fact a true hypothesis. A hypothesis that combines the true functional form and parameters $\beta = 3$ and $\tau = 1$ would be a false hypothesis. Hypotheses can also include other functional forms.

Size is calculated from the proportion of true hypotheses rejected in step 4 at the range of $\alpha$ levels.

Similarly, power is calculated as the percentage of false hypotheses rejected at a given level. For these simulations, we use an $\alpha = 0.05$. We report these rejection rates in the following simulations.

*5.1.1 Test Statistics*

Testing requires test statistics. A useful test statistic should be small when models do a good job of aligning the treated and control groups. When models do a poor job, the value of the test statistic should be large. This property is known as "effect increasing" (Rosenbaum, 2002, Proposition 2, Chap 2). With respect to the parameters in the model, we want to find a test statistic that is effect increasing on both $\beta$ and $\tau$. The first test statistic, a mean difference statistic, compares the means of the treated and control groups. The second test statistics, the KS test, takes the supremum of the absolute difference between the empirical CDFs of the control and treated distributions. The final test statistic, the Mann-Whitney U, is the normalized sum of the ranks of the treated units. We selected these three test statistics to represent a range of trade offs available to researchers. Differences of means are very efficient when data are Normal, but can lack power on non-Normal data. Like the difference of means, the Mann-Whitney U statistic is a location based statistic, but retains good power on non-Normal data (Keele, McConnaughy and White, 2012). The final test statistic, the Kolmogorov-Smirnov (KS) test statistic represents a different approach in that it can detect more general divergences between the treated and control distributions that include but are not limited to means or medians (or other specific locations).

Figure 8 shows how well three potential test statistics perform as we vary $\beta$ and $\tau$ away from the true values used in our simulation. For this simulation, we use a sample size of 256 units, with 512 edges (i.e. the network shown in Figure 1) and a uniformity trial where baseline outcomes for unit $i$ depend on the density of the network connections for that unit: using the function $f(x) = \beta + (1-\beta)\exp(-\tau^2 x)$, our growth curve equation, we set $\mathbf{y_0} = U(30, 70) \cdot f(\mathbf{1}^T \mathbf{S})$. The plot labeled "Size" shows that all three test statistics maintain appropriate size when the null hypothesis is true. That is, if we are willing to tolerate an error rate of 5%, this test rejects falsely no more than 5% of the time (that this is true for all $\alpha$ is shown by the clustering of the distribution right along and below the 45 degree line and within the dashed lines showing the ±2 bounds of the standard error of simulation). The plots labeled "Power" show that all three statistics have similar power as $\beta$ varies away from the truth: that is, as we test hypotheses about $\beta$ which are further from $\beta = 2$, more and more of our tests will produce low $p$-values (indicating a lack of congruence between our hypotheses and the data) until almost all tests of certain extreme hypotheses will produce $p$-values less than .05 (power of 1.0 for $\alpha = .05$). With respect to $\tau$, all three statistics again perform similarly. We omit simulation error bars on the power analyses plots to avoid clutter. We use the KS statistic in all the subsequent simulations because differences between treatment and control groups arising from hypotheses about interference are likely to cause more than simple shifts in the location of the two groups and so test statistics sensitive only to differences in location will tend to miss differences in spread, skew, kurtosis that might indicate useful model.

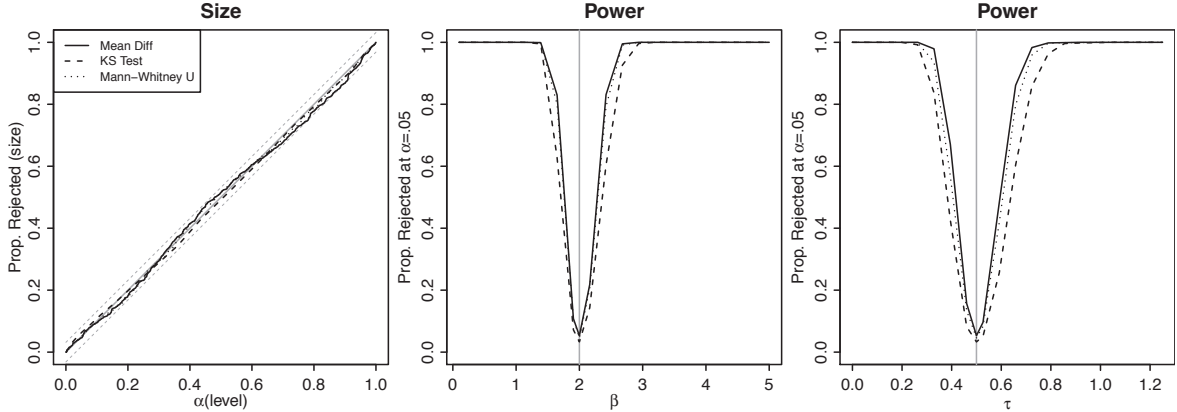

Figure 8: Simulations for the true spillover model tested with three different test statistic. The three statistics are a mean difference statistic, the Mann-Whitney U (rank) statistic, and the Kolmogorov-Smirnov statistic. For $\beta$-power simulations, $\tau = 0.5$. For $\tau$-power simulations, $\beta = 2$. Dashed lines on the "Size" panel show ±2 standard errors of simulation where $SE_{\text{sim}} = \sqrt{p(1-p)/n}$ (Imbens and Rubin, 2009, Chap 5).

### 5.1.2 Baseline Outcomes

In this simulation, we vary the uniformity trial used to generate the data (i.e., the result that would be observed if all units received the control condition). We consider three kinds of fixed uniformity trial data. In the first case, we generate the uniformity trial data by sampling uniformly from 30 to 70: $\mathbf{y_0} = U(30, 70)$. We label these data the "base" uniformity trial data. For the second uniformity trial vector, we multiply the base uniformity data by the maximum possible spillover for each unit using the function $f(x) = \beta + (1-\beta)\exp\left(-\tau^2 x\right)$: $\mathbf{y_0} = U(30, 70) \cdot f(\mathbf{1}^T \mathbf{S})$. While the first uniformity data set is completely independent of the network, in the second data set, units with many connections will have systematically higher uniformity trial data. These data can be thought to be "pre-dosed," in the sense that each unit has received the equivalent effect as if all her neighbors had been treated. All $y_{i,0}$ in the second trial are somewhere between 1 and $\beta$ times as large as the first uniformity data set. As the network multiplies outcomes, we label these data "network plus." For the last data set, we again use the base data, this time dividing by the function $f$: $\mathbf{y_0} = U(30, 70)/f(\mathbf{1}^T \mathbf{S})$. We label these data "network minus."

Figure 9 shows how the test performs under these different initial conditions. For each uniformity trial data, we create 1000 simulated data sets using our spillover model and parameters $\beta = 2$ and $\tau = 0.5$. For each data set, we apply the algorithm given in § 4.1 using the KS test statistic. All three uniformity trials are comparable with respect to $\beta$. With respect to the $\tau$ parameter, the pre-dosed "network plus" makes rejecting false hypotheses somewhat harder, though power approaches one within a factor of two to the true $\tau$ parameter. All three approaches maintain appropriate size within expected simulation error (grey bars). In an effort to make our simulations more realistic, we select the "network plus" distribution for the uniformity trial in all the other simulations in this section.[15]

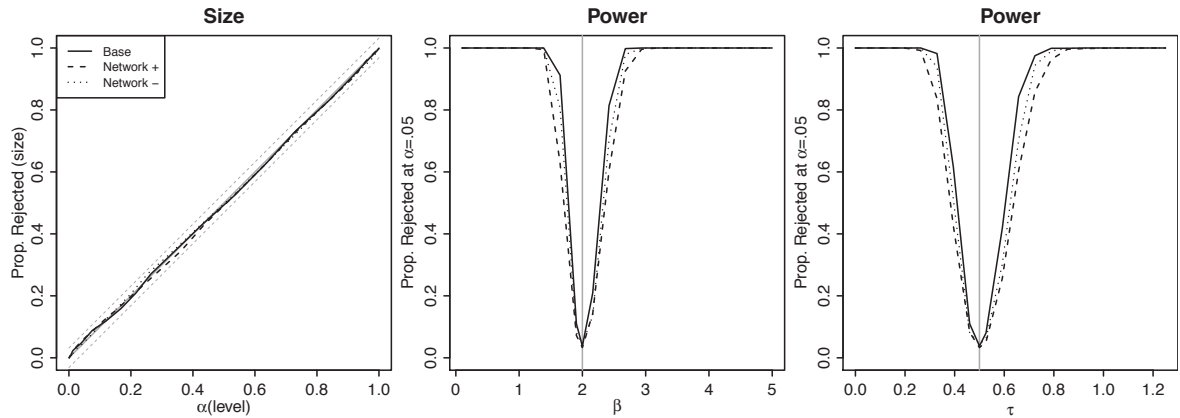

Figure 9: Uniformity trial simulations for the true spillover model. For $\beta$-power simulations, $\tau = 0.5$. For $\tau$-power simulations, $\beta = 2$.

These simulation results indicate that the distribution of the baseline outcomes matters to the test's ability to discriminate between hypotheses. While researchers will never observe the uniformity trial as would be experienced by subjects in the actual experiment, there are opportunities to learn about the uniformity trial from existing data on baseline outcomes. For example, in a turnout experiment, it is likely the uniformity trial would be similar to historical turnout data. In an experiment conducted on a panel subject pool, pre-tests or previous experimental results can serve to suggest a distribution for the uniformity trial.

---

[15]It was also the uniformity trial used in generating the data in section 3.

### 5.1.3 Sample Size

In this simulation we vary the sample size from small to large with sample sizes of 32, 256, and 1024. At each sample size, we fix the number of edges in the network equal to $2n$. The 1000 observed data sets are generated using the "network plus" uniformity trial data and the standard $\beta = 2$ and $\tau = 0.5$ values. The models are assessed using the KS test statistic. Figure 10 shows the size of the test and the power to reject false hypotheses across a range of false $\tau$ and $\beta$ values.

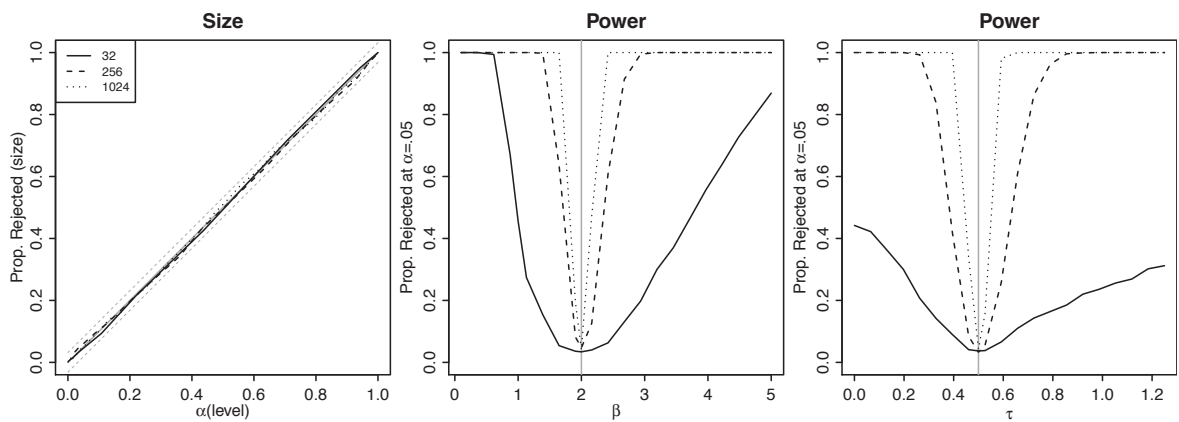

Figure 10: Sample size simulations for the true spillover model. For $\beta$-power simulations, $\tau = 0.5$. For $\tau$-power simulations, $\beta = 2$.

As expected, a small sample size of 32 units has difficulty rejecting hypotheses close to the truth. It takes roughly a four fold increase or decrease in the hypothesized value of $\beta$ before the 32 unit sample can reject false hypotheses at the $\alpha = 0.05$ level. The small sample size lacks the power to reject almost any tested value of $\tau$ at this level. Larger sample sizes, however, do enable rejection of false hypotheses even when these hypotheses are quite close to the true $\beta$, and power increases as sample size increases. Sample size matters slightly more in discriminating for spillover hypotheses, related to $\tau$ at least for this particular model. Even for small sample sizes, test size remains within simulation error, and tends to be slightly conservative, failing to reject the true null more often than necessary as given in the $\alpha$-level ($x$-axis).

These plots, unsurprisingly, suggest that larger sample sizes are always better for increasing power. At the same time, even small sample sizes appropriately limit type I error, so researchers are unlikely to be misled into rejecting true hypotheses too often. Researchers will have to balance the costs of increasing sample sizes against the benefits of additional power. At least for this simulation setup, if a researcher were primarily interested in the direct effect of $\beta$, the largest sample size does not confer much additional power over a sample one quarter the size. If the spillover parameter is more important, the large sample size premium may be worth the additional cost depending on expected effect size.

*5.1.4 Network Density*

More dense networks may have either more, or less, information available for assessing hypotheses, depending on the model at hand. For example, very dense networks may provide excellent information about spillover effects but enable us to learn very little about direct treatment effects. In this simulation, we vary the number of edges in the network, the network density, to learn about our model and data. To vary the network density, we fix the sample size at 256 units and investigate the test properties when there are $n \cdot (0.25, 1, 2, 5)$ edges in the network. Edges are added by finding the closest pair, adding an edge between the pair, and repeating the process until enough edges are in the network. The uniformity trial is the "network plus" uniformity trial, created for the 256 unit, 512 edge network. We use the KS test statistic as before. Figure 11 shows that the test continues to maintain the appropriate size, as specified in the $\alpha$-level. The power plots display several interesting results. First, while having zero edges in the network provides great power against false hypotheses over $\beta$, we are unable to say anything about $\tau$. This result makes sense: When there are no edges in the network, there is no observed spillover. Therefore, the experiment provides no insight into the true value of $\tau$. *All* hypotheses about $\tau$ imply the same adjustment to the data and therefore have the same test statistic value. Consequently, the design provides no useful information about $\tau$ and, correctly, would produce infinite confidence intervals. We will see this result again in § 5.2, when we use a data set with a network but with data that are generated with a true $\tau = 0$ parameter.

Conversely in the power plots, having a dense network (i.e., having $256 \times 5 = 1280$ edges) diminishes power for $\beta$ but increases power for $\tau$. In fact, there appears to be a minimum density required, at least for this data set, somewhere around 1, for maintaining power against reasonable, but false, values of $\tau$. Researchers who control networks (such as those in laboratories) may desire to trade sample size for network characteristics like density to achieve optimal power against both direct and indirect parameters depending on the results of simulations like these using their own theoretical causal models.

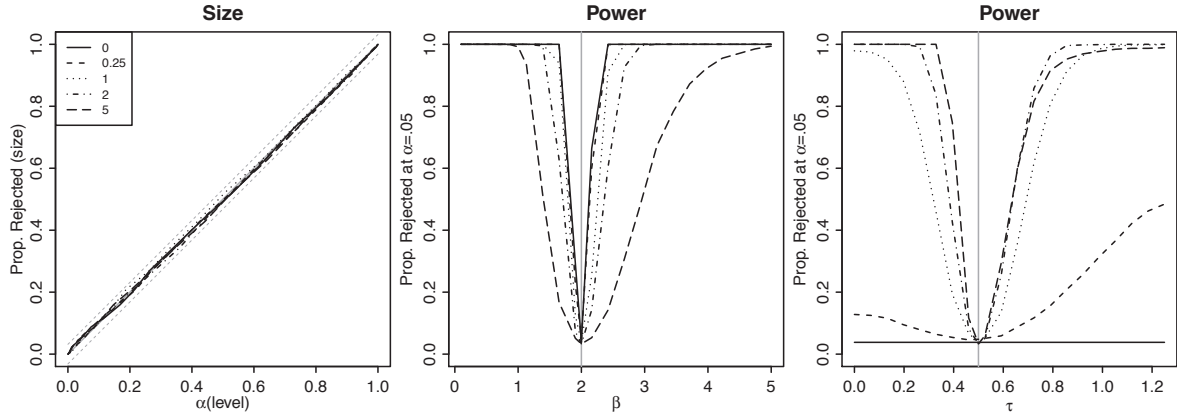

Figure 11: Network density simulations for the true spillover model. For $\beta$-power simulations, $\tau = 0.5$. For $\tau$-power simulations, $\beta = 2$.

### 5.1.5 Percent Treated

To vary the percent treated, we fix the sample size at 256 and the number of edges at 512. As usual, we use the "network plus" uniformity trial and the KS test statistic. We then test hypotheses when the number of treated units in the simulation is 10%, 25%, 50% and 75% of the sample size. Figure 12 shows these results.

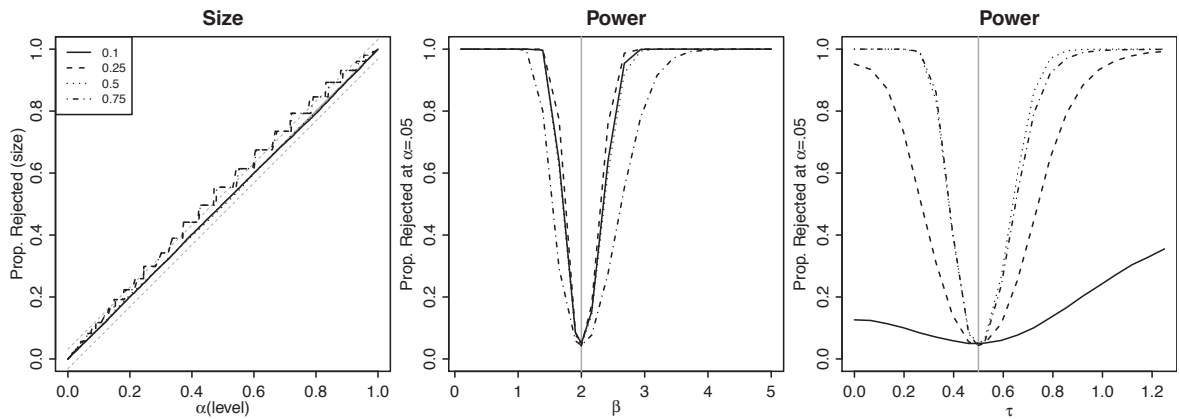

Figure 12: Percent of subjects that are assigned to treatment simulations for the true spillover model. For $\beta$-power simulations, $\tau = 0.5$. For $\tau$-power simulations, $\beta = 2$.

For our network and model, the most powerful design involves randomly assigning half of the units to treatment and half to control. Note, however, that the designs over-sampling and under-sampling treated units do not perform equally with respect to rejecting large, false values of $\tau$. While assigning 75% of the subject pool to the treated condition performs almost as well as assigning 50%, assigning only 25% of the subjects to treatment performs much worse across the entire plotted range. For other datasets, networks, and models, these results may even be exacerbated to the point that over or under-sampling strictly dominates equal numbers of treated and control units. While not shown here, we have seen this result in other simulations. The exact nature of these curves depends strongly on the underlying data, network, and model, so we strongly encourage researchers to perform similar power simulations before deciding how to allocate treatment or how to interpret the *p*-values from their tests.

### 5.1.6 Simultaneous Power Analysis of $\beta$ and $\tau$

In the previous simulations, in the name of clarity, we have presented separate power analysis for $\beta$ and $\tau$. For the $\beta$ power analyses, we have held $\tau$ fixed at the true value of $\tau = 0.5$. For the $\tau$ power analyses, we have held $\beta$ fixed at $\beta = 2$. In this simulation, we vary both parameters simultaneously and display the results. We again use the 256 unit network with 512 edges, with outcome generated from applying the true model to the "network plus" uniformity trial. As before, we employ the KS test statistic. Figure 13d shows the size of the test, which shows that the test continues to reject the true joint null at appropriate rates. Moreover, looking across all repetitions of simulation, the true value is within the 2-dimensional 95% confidence region 96.7 percent of the time, within expected simulation error.

Figure 13a shows the proportion of hypotheses rejected as hypotheses over both $\beta$ and $\tau$ vary: Joint hypotheses that diverge from the truth are rejected nearly always for most of the plot. Yet, rejection rates decrease from 100% for hypotheses with large direct and large spillover effects, an intuitive result. Large amounts of spillover can mask direct treatment effects in this model, so this simulation teaches about our model: these hypotheses are observationally very similar to the true hypothesis. The lines bisecting the plot correspond to the true values of $\beta$ and $\tau$. Subplot 13b shows the power along the line $\tau = 0.5$, while subplot 13c shows the power along the line $\beta = 2$. These plots correspond to the power plots in the previous simulations.

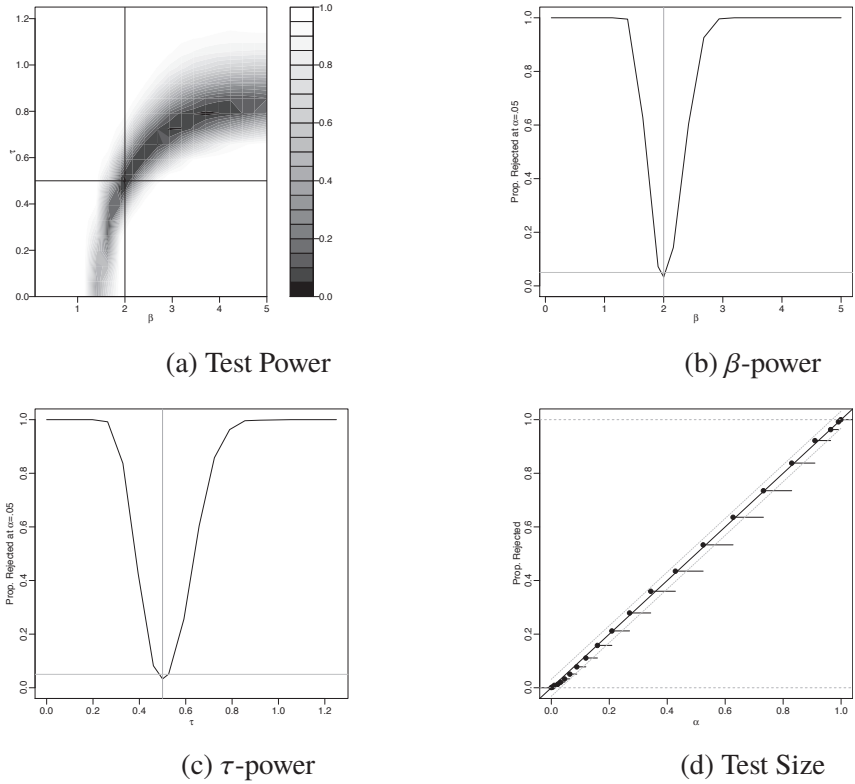

(a) Test Power

(b) $\beta$-power

(c) $\tau$-power

(d) Test Size

Figure 13: Size (rejection of true hypothesis at given level) and power (rejection rate of false hypotheses) plots for a single model investigation. Horizontal gray lines in panels (b) and (c) show that 5% of hypotheses about the true values are rejected at the $\alpha = .05$ level.

## 5.2 Comparing Models

Recall that to reject a hypothesis, the adjustment implied by the model $\mathcal{H}(\mathbf{y_z}, \mathbf{0}, \theta) = \mathbf{y_0}$, the function that turns observed data into the uniformity trial, must produce large differences in the treated and control groups. Conversely, a hypothesis that we fail to reject will do a good job of making the treated and control groups appear similar. One method for assessing models in this way was suggested by Rosenbaum (2010, Chap 2, p.48) in which distributions of outcomes after and before application of models were displayed using boxplots. Figure 14 shows the distribution of the observed data and the distribution of the model adjusted data for three hypotheses: the true data generating process, a hypothesis for which we have low power in Figure 13, and a hypothesis that we would reject at a high rate from that same figure. Since we are in the position to know the uniformity trial data, we display that as well. Both the true hypothesis and the false, but low power, hypothesis clearly align the data better than the high power hypothesis, with the true hypothesis aligning the distributions more closely. Of course, since we know the uniformity trial, we can see that the true model restores the treated and control groups to be, what are in essence, random samples from the uniformity trial. In practice we do not have access to this knowledge. As Rosenbaum (2007) points out, we cannot distinguish between models that imply similar effects to both treated and control groups. In the low power hypothesis, increased spillover and a larger main effect makes the treated and control groups appear very similar. Even with more nuanced models of spillover effects, there is a price to be paid when spillover occurs: it is increasingly difficult to discriminate between competing models. Yet, the hypothesis against which we have high power keeps the two distributions fairly distant, even if closer than the observed difference.

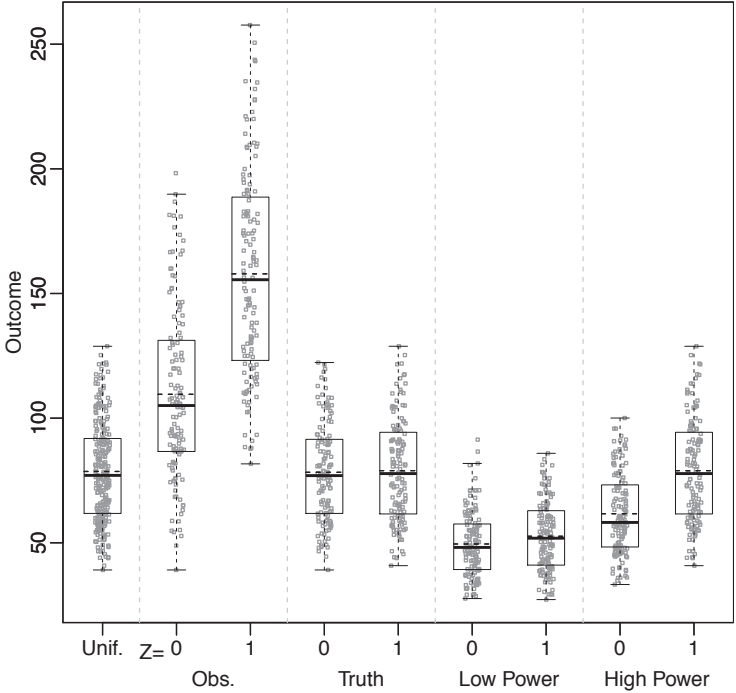

Figure 14: Comparison of models: the true model ($\tau = 0.5, \beta = 2$), an infrequently rejected false hypothesis ($\tau = 0.75$ and $\beta = 3$), and a frequently rejected false hypothesis ($\tau = 1$ and $\beta = 2$). Data are grouped by treatment ($Z = 1$) and control ($Z = 0$). Note also that the observed data are equivalent to the adjustment implied by the sharp null of no effects.

The previous simulations suggest that the methodology presented here does fulfill the minimal criteria that we tend to expect from our statistical procedures: test size is less than or equal to the level of the test and the test has good power against false sets of parameters. Both of these properties derive from the model being true. What about when the model is false?

We begin to address this question by starting from the simplest alternative true model: the sharp null of no effects. This model states that the intervention had no effect on the population. In other words, the observed data are precisely what would have been observed if no intervention had been made. In fact, the sharp null of no effects is a special case of the model in equation 6 when $\beta = 1$. Many parameterized models exhibit this trait. For the additive, non-interference model $\mathcal{H}(y_i, \mathbf{z}, \mathbf{w}, \alpha) = y_{i,\mathbf{z}} - \alpha(z_i - w_i)$, the sharp null occurs at $\alpha = 0$. Testing the sharp null as parameterized for one model is equivalent to testing the sharp null as parameterized for a different model as the adjustment of the data is the same. Therefore, if we reject or fail to reject the sharp null for one functional form, we need not try a different functional form. The sharp null will be rejected or not for the new model as well.

For some models with multiple parameters, only one of the parameters will represent the sharp null hypothesis. In the case of the spillover model of Equation 6, when there is no direct effect, no spillover effects are possible. When $\beta = 1$, *all* values of $\tau$ imply the same adjustment to the data. To demonstrate this property, we simulate 1000 experiments in which treatment had no effect using our standard 256 unit, 512 edge subject pool and based on the "network plus" uniformity trial. As the power plot in Figure 15 shows, the test almost always fails to reject the true model (the line $\beta = 1$), the sharp null of no effects. Looking across all simulations and using an $\alpha$-level of 0.05, for all values $\tau$, hypotheses that include $\beta = 1$ are rejected in 3.3 percent of the simulations, within simulation error. If the intervention had no effect, our method protects researchers from rejecting this null too often.

We next consider the case where the direct effect is correct, but there are no spillover effects. Specifically, we test the model where $\beta = 2$ and $\tau = 0$. Figure 16 shows the now standard power graph for the simulation. As before, the true model is infrequently rejected, as expected. Moreover, the range of plausible hypotheses for which spillover exists is small. A researcher would be unlikely to be misled into thinking spillover occurred in large amounts if none in fact existed.

In our final simulations, we consider the performance of the test when the functional form of the model is misspecified. In the next simulation, instead of generating data from the multiplicative model of Equation 6, we generate data for this simulation using a simple additive model:

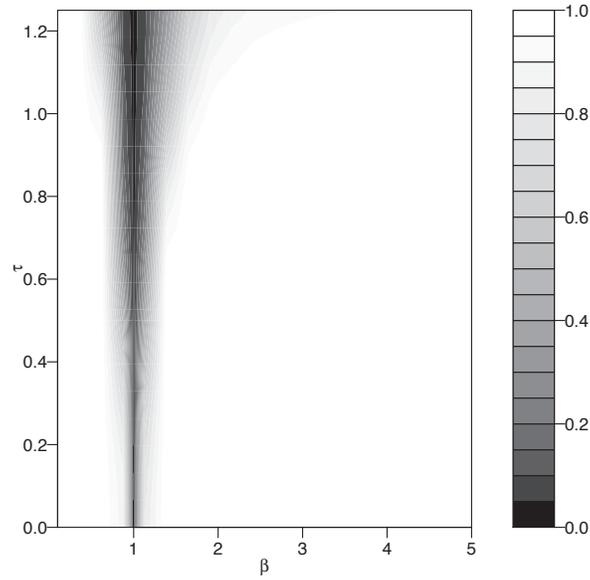

Figure 15: Proportion of $p$-values below .05 for joint hypotheses about $\beta$ and $\tau$ as defined in equation 6 when the true model is the sharp null model (setting $\beta = 1$ in the simulation engine as described in § 5.1.6.

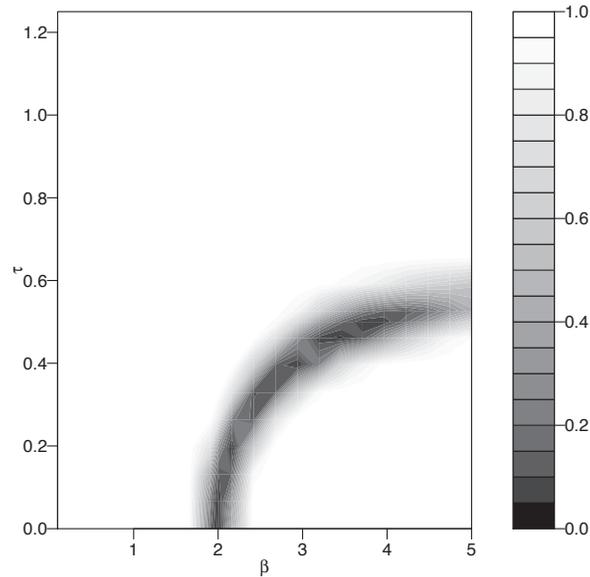

Figure 16: No spillover data ($\tau = 0$), with a main effect of $\beta = 2$. The data are 256 units, with 512 edges using the "network plus" uniformity trial.

$$\mathcal{H}(y_i, \mathbf{z}, \mathbf{w}, \alpha) = y_{i,\mathbf{z}} - \alpha(z_i - w_i) \qquad (8)$$

For each of the 1000 repetitions, we generate outcomes consistent with Equation 8 with a true $\alpha = 48$. This value was chosen as it is the average difference between the treated and control groups in the data used in Section 2; therefore, both models set the treated and control group means to similar values. We continue to use the "network plus" uniformity trial and the KS test statistic.

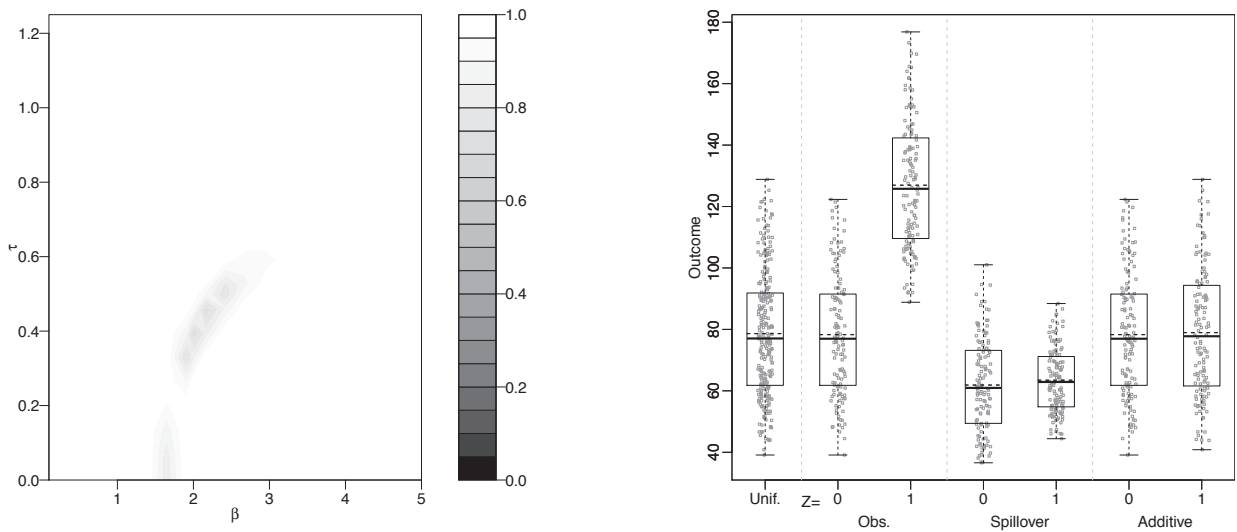

(a) Proportion of hypotheses rejected for hypotheses generated by the spillover model applied to data generated by the additive model. Minimum achieved at $\beta = 2$ and $\tau = 0.395$.

(b) Example observed data (from one of the 1000 trials) and adjustments implied by the true additive model and the spillover model at $\beta = 2$ and $\tau = 0.395$.

Figure 17: Simulation results for data generated by the true additive model (Eq. 8, $\alpha = 48$) and tested using the spillover model (Eq. 6).

Figure 17a shows the rate of rejection (at the $\alpha = 0.05$ level) for hypotheses generated using the multiplicative spillover model of Equation 6. Almost all of these hypotheses are rejected in almost all of the 1000 tests. The least rejected hypothesis is $\beta = 2$ and $\tau = 0.395$, which is rejected in 71.4% of the simulations. As intended, this hypothesis is close to the true values we used in the previous simulations; however, the rejection rate is much, much higher. For these data, the test maintains good power against incorrectly specified models, rejecting false hypotheses at a high rate. Yet, rejecting so often is a sign that the model is quite incongruent with the data.

In Figure 18a we put the shoe on the other foot. We generate data consistent with the spillover model and test using the additive model. In this plot, we see the least rejected hypothesis is $\alpha = 48.718$, which is rejected in 10.8% of the simulations. As noted previously, this is roughly the value of $\alpha$ that most closely mimics the true values of $\beta$ and $\tau$.

In these two simulations, we see that incorrect functional forms are not always rejected. When

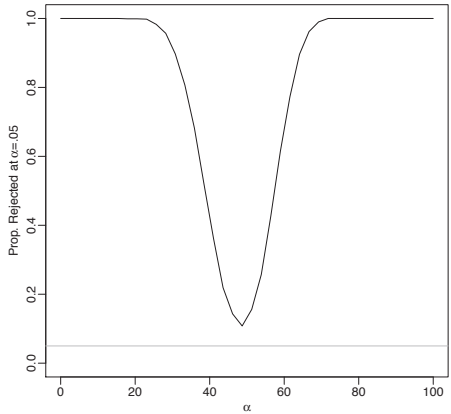 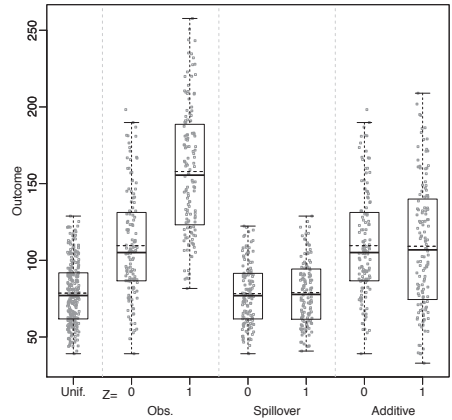

(a) Proportion of hypotheses rejected for hypotheses generated by the additive model applied to data generated by the spillover model. Minimum achieved at causal parameter $\alpha = 48.718$. Horizontal line at 0.05.

(b) Example observed data (from one of the 1000 trials) and adjustments implied by the true spillover model and the additive model at $\alpha = 48.718$.

Figure 18: Simulation results for data generated by the true spillover model (Eq. 6, $\beta = 2$ and $\tau = 0.5$) and tested using the additive model (Eq. 8).

data are generated from one process, it can appear similar to data generated by another process. Figures 17b and 18b provide some insight into how these two models are similar. Both plots—one for each of the previous two simulations—show the uniformity trial data and the observed data generated by the true model. For Figure 17b, the additive model (at the true parameter) aligns the data very well. The spillover model adjustment, while not perfectly aligned, is still reasonably close. Misalignment would be achieved in 71.4% of experiments, just due to chance, even if the spillover model were true. Likewise, in Figure 18b, the misalignment created by the additive model adjustment would occur in 10.8% of experiments, just due to chance. These plots illustrate an important point: our statistical methods are not oracles. They can only tell us if the model's perspective on the data is implausible. For a given data set, many hypotheses are likely. This testing framework can only help us eliminate implausible hypotheses, not accept plausible models.

## 5.3 Summary of Simulation Studies

Different researchers will have different models arising from different theories, and face diverse designs and datasets. Our aim in this section has been to suggest some approaches to model and test assessment while also answering some important questions about a new method. We show that our method meets the standards commonly required of statistical tests: it is never overly likely to encourage us to reject a true null hypothesis, and, in a wide variety of situations, it has power to reject hypotheses that are not true—and such power is a function, as one would hope, of the amount of relevant information used in the testing procedure. Here, relevant information includes sample size, network density, baseline outcomes, relationships with the network, and proportion treated. Test statistics use information differently, and we also assessed differences between the operating characteristics of a few test statistics. Even with increases in power, data may be consistent with more than one model. Graphical methods provide insight on the adjustments implied by models. The studies that we have shown here will be useful both for the design of studies but also for learning about the ways that complex models may interact with complex designs and data.

## 6 Discussion

When treatments given to one unit can change the potential outcomes for another unit, the consequences of ignoring interference may be serious. Imagine a development project aiming to assess a policy applied to different villages in need of aid. If members of control villages communicate with members of treated villages, then scholars will have trouble advising policy makers about whether the policy should be rolled out at a large scale. We have long known, in fact, that the average treatment effect is not even well identified or meaningful under interference (Cox, 1958).

So far attempts to enable statistical inference about treatment effects with interference have taken for granted the average treatment effect framework and worked to partition the average into parts attributable to interference and parts attributable to direct experience with the treatment. In this paper, we propose a different approach based on asking direct questions about specific forms of interference. Fisher's test of the sharp null is still meaningful even when each unit may have many potential outcomes due to interference. Additionally, Fisher's framework allows detection of interference (Aronow, 2012), and under certain conditions, allows the creation of intervals for hypotheses about treatment effects without requiring specific statements about the form of interference (Rosenbaum, 2007). Our paper contributes to this literature by showing how one may directly specify and assess hypotheses about theorized forms of interference. We also show that one may present and summarize information that illuminates the information contained in a dataset regarding different combinations of hypotheses about interference and treatment effects. This form of statistical inference does not require asymptotic justifications or assumptions about the stochastic processes generating outcomes although fast and reliable approximations are available as samples grow in size.

While not required, such additional assumptions can profitably speed computation. For example, in our simulation, we employed several test statistics, all of which have convenient large sample approximations. While the asymptotic justification would have been inappropriate for a small sample, in larger samples it can be quite useful. For small samples, the enumeration scheme suggested in the § 4.1 provides exact $p$-values without requiring any appeals to asymptotic results. As computational power increases, researchers may prefer exact solutions even when approximations exist. This framework is flexible enough to accommodate both approaches. A benefit of our simulation based approach to assessing the method is that one would detect whether the large sample approximations were a problem (for example, by noticing that the Type I error rate was not controlled).

The software used in this analysis is not difficult to use. In Appendix A we provide several example code snippets to illustrate the simple relationship between a formal model of effects and a statistical test of relevant hypotheses. If researchers can write down a model of effects mathematically, they can use the software involved. Appendix A shows how the model used in this paper can be implemented, which we hope serves to show how straight-forward the process is. We continue to extend and simplify the software, especially in the area of generating simulations such as those used in this paper.

In this paper, we have allowed theory to be prior to selection and application of a statistical technique. As we have portrayed our method, researchers first select a theory or set of theories, and then write models that capture the implications of those theories for subjects in an experiment, including spillover effects. We hope that our method also encourages researchers to generate and engage with new theories and models. By adding another tool to their toolbox, researchers now have a new language to transcribe theories, opening up new avenues for theoretical advances. We are especially optimistic about the opportunities for assessing models arising from formal theories using the techniques in this paper. We think that our framework is especially well suited to assess actor level models in which actors observe the treatment status of others and anticipate the actions of their neighbors. Connecting the Fisherian method of inference to game theoretic models and concepts is an area ripe for exploration.

Our proposed method focuses attention on unit level models: models that explain how individual subjects take on values in our experiments. The natural evaluation tool for models is the hypothesis test, which also has a central role in our proposal. Excellent research also exists on estimating effects, specifically average treatment effects in the presence of spillover (see § 1 for an extensive list). We consider it an opportunity for future research to connect average effect estimation with the unit level, testing based framework suggested in this paper. For example, randomization-based approaches to the estimation of average treatment effects (Aronow, 2012; Tchetgen and VanderWeele, 2010) and testing weak null hypotheses (Rosenblum and Van Der Laan, 2009) may provide both computational advances and conceptual advances when we think about them from the perspective of the sharp null hypothesis. This set of approaches have similar goals and serve complementary purposes. One might imagine a workflow in which average direct and indirect effects are reported along with tests of the weak null and models of unit level processes leading to those effects are assessed. We suggest that our method is especially useful to researchers who have a clear substantive theory as to how spillover occurs. By writing down unit level models, these theories enter the statistical analysis very directly.

Additional work is needed in categorizing, describing, and diagnosing models. For example, different models may exhibit different parameter sensitivity (or lack effect increasing characteristics). An ability to describe the features of a model, even before data have been collected, appears useful. When models contain more than two parameters, we will have to consider how to present and summarize such results, as the 1 and 2 dimensional plots in this paper would be insufficient. As we noted in our simulation studies, models sometimes imply similar adjustments to data. Precisely describing the relationships between models is an open topic. We encourage researchers to engage with these issues as they apply the techniques in this paper in their own work. It is clear that we have just scratched the surface of the of the world of theoretically driven interference models.

# Appendix A  Code Examples

This appendix provides a brief behind the scenes look at how the simulations in § 5 were implemented. The simulations rely heavily on the software package RItools . This appendix is a useful demonstration of how an applied researcher might begin an experiment by first simulating outcomes and trying different design elements (e.g., sample sizes) and models. For a complete picture of how our simulations were generated, see the source code to this entire paper at [URL].

## Appendix A.1  Model

Our model here is based on a particular network described in an adjacency matrix $\mathbf{S}$. To generate the model function, we parameterize on $\mathbf{S}$ and return a `UniformityModel` object, which is an amalgam of two functions. The first maps observed data to the uniformity trial $\mathcal{H}(\mathbf{y_z}, \mathbf{0}, \beta, \tau)$. The second maps uniformity trial data to what would be observed under the model $\mathcal{H}(\mathbf{y_0}, \mathbf{z}, \beta, \tau)$. These functions are parameterized on $\beta$ and $\tau$ as discussed in § 5.

```
## The UniformityModel is a datatype defined in the R package on CRAN.
## The first argument is a function to go from observed data to the uniformity trial
## The second argument is the inverse of the first model (it generates
## observed data from the uniformity trial)

## y0 is the uniformity trial
## y is the observed outcome
## z is 0 or 1 denoting treatment assignment

## Since this model depends on S, we need to generate for a specific S
## The growthCurve function specifies how spillover happens.

growthCurve <- function(beta, tau, x) {
  (beta + (1 - beta) * exp(-(tau^2) * x))
}

interference.model.maker <- function(S) {
  ## just to be safe about lazy argument evaluation,
  ## probably not needed
  force(S)

  ## we will return a UniformityModel object
  UniformityModel(
    function(y, z, beta, tau) {
      zS <- as.vector(z %*% S)
      z * (1 / beta) * y +
        (1 - z) * y / growthCurve(beta, tau, zS)
    },
    function(y0, z, beta, tau) {
      zS <- as.vector(z %*% S)
      z * beta * y0 +
        (1 - z) * growthCurve(beta, tau, zS) * y0
    })
}
```

## Appendix A.2  Sample Size Simulation

We include here the sample size simulation as an illustration. The general flow of this simulation (and the other simulations as well) is as follows:

1. Create the uniformity trial data for each sample size.

2. For each sample size:

    (a) Draw 1000 treatment assignments consistent with the design.

(b) Generate 1000 observed datasets using the uniformity trial data and the treatment assignments from the previous step based on the true model.

(c) Run `RItest` on a search space of parameter values. These parameter values include the true model.

The `RItest` function handles the statistical inference algorithm, reporting a $p$-value for each parameter value tested. By looking across these results, we get a picture of how often the true parameter value is rejected at a given $\alpha$-level (type I error) and how often false parameter values are rejected (power). The last part of the simulation does a little processing to make these error rates easier to display and report.

```r
# Read in the constants (like DENSITY, REPETITIONS, SEARCH, TRUTH) and
# functions like simulationData(), and load relevant packages (such
# as the package containing RItest() and invertModel() and sampler(), etc.
source("simulation/setup.R")

# Simulation 1: Increasing sample size.
sampleSizes <- c(32, 256, 1024)
edges <- floor(sampleSizes * DENSITY) # fix the number of edges per sim

## For each sample size make a dataset with baseline data depending on
## and adjacency matrix, S, which, in turn, depends on DENSITY.
sampleSizeData <- vector("list", length(sampleSizes))
for (i in 1:length(sampleSizes)) {
  sampleSizeData[[i]] <- simulationData(sampleSizes[i], edges[i])
}

runSim<-function(d,SEARCH){
  ## create data consistent with the interference model
  data <- d$data
  n <- dim(data)[1]
  nt <- n/2 # also equals nc in this simulation
  sampler <- simpleRandomSampler(total = n, treated = nt)

  ## get the particular model given this network
  model <- interference.model.maker(d$S)

  Zs <- sampler(REPETITIONS)$samples # discard the weight element

  res <- apply(Zs, 2, function(z) {
    y <- invertModel(model, data$y0, z, TRUTH$beta, TRUTH$tau)
    RItest(y, z, ksTestStatistic, model, SEARCH, type = "asymptotic")
  })

  return(res)
}

# For each element of simOneData run RItest for a set of values around
# the truth. Here, we speed computation by assessing each parameter
# separately. We could search the 2d grid by simply using SEARCH
# (defined in setup.R) rather than SEARCH.TAU and SEARCH.BETA. RItest
# will search any size grid.

sampleSizeTauResults <- lapply(sampleSizeData, function(d){
  runSim(d,SEARCH.TAU) })

sampleSizeTauPower <- simulationPower(sampleSizeTauResults)

sampleSizeBetaResults <- lapply(sampleSizeData, function(d) {
  runSim(d,SEARCH.BETA)})

sampleSizeBetaPower <- simulationPower(sampleSizeBetaResults)

names(sampleSizeTauPower) <- names(sampleSizeTauResults) <-
```

```
  names(sampleSizeBetaResults) <- names(sampleSizeBetaPower) <- names(sampleSizeData) <-
      sampleSizes

save(file = "simulation/samplesize.rda", sampleSizeTauResults, sampleSizeTauPower,
      sampleSizeBetaResults, sampleSizeBetaPower, sampleSizeData)
```

**Appendix A.3  Testing Hypotheses from a Model**

In the preceding simulation hypotheses were tested many times. To assess one set of hypotheses as generated by one model (for example, as shown in Figure 6) one would do the following (although one would not have to create data for use in testing in a real application):

```
source("simulation/setup.R")

## Create the canonical dataset
canonicalData <- simulationData(256, 512)
canonicalZ <- as.vector(canonicalData$sampler(1)$samples)
canonicalModel <- interference.model.maker(canonicalData$S)
canonicalOutcome <- invertModel(canonicalModel,
                                canonicalData$data$y0,
                                canonicalZ,
                                beta = TRUTH$beta,
                                tau = TRUTH$tau)

## Test joint hypotheses generated from the model
canonicalRI <- RItest(canonicalOutcome,
                      canonicalZ,
                      ksTestStatistic,
                      moe = canonicalModel,
                      parameters = SEARCH,
                      type = "asymptotic")
```

# References

Aronow, Peter M. 2012. "A General Method for Detecting Interference Between Units in Randomized Experiments." *Sociological Methods & Research* 41(3):3–16.

Aronow, Peter M. and Cyrus Samii. 2012*a*. "Estimating Average Causal Effects Under General Interference.".

Aronow, Peter M. and Cyrus Samii. 2012*b*. "Large Sample Coverage Properties of Inverted Exact Test Intervals for Randomized Experiments.".

Barnard, GA. 1947. "Significance tests for 2× 2 tables." *Biometrika* 34(1/2):123–138.

Berger, R.L. and D.D. Boos. 1994. "P Values Maximized over a Confidence Set for the Nuisance Parameter." *Journal of the American Statistical Association* 89(427).

Brady, Henry E. 2008. "Causation and explanation in social science." *Oxford handbook of political methodology* pp. 217–270.

Chen, J., M. Humphreys and V. Modi. 2010. "Technology Diffusion and Social Networks: Evidence from a Field Experiment in Uganda.".

Cox, David R. 1958. *The Planning of Experiments*. John Wiley.

Fisher, R.A. 1935. *The design of experiments. 1935*. Edinburgh: Oliver and Boyd.

Hansen, Ben B. and Jake Bowers. 2008. "Covariate balance in simple, stratified and clustered comparative studies." *Statistical Science* 23(2):219–236.

Hansen, Ben B. and Jake Bowers. 2009. "Attributing Effects to A Cluster Randomized Get-Out-The-Vote Campaign." *Journal of the American Statistical Association* 104(487):873—885.

Hollander, Myles. 1999. *Nonparametric Statistical Methods, 2nd Edition*. Second ed. Wiley-Interscience.

Hong, G. and S.W. Raudenbush. 2006. "Evaluating Kindergarten Retention Policy." *Journal of the American Statistical Association* 101(475):901–910.

Hudgens, M.G. and M.E. Halloran. 2008. "Toward causal inference with interference." *Journal of the American Statistical Association* 103(482):832–842.

Ichino, N. and M. Schündeln. 2011. Deterring or Displacing Electoral Irregularities? Spillover Effects of Observers in a Randomized Field Experiment in Ghana. Technical report Working paper.

Ichino, N. and M. Schündeln. 2012. "Deterring or Displacing Electoral Irregularities? Spillover Effects of Observers in a Randomized Field Experiment in Ghana." *Journal of Politics* 74(1):292–307.

Imbens, G. and D. Rubin. 2009. "Causal Inference in Statistics." Unpublished book manuscript. Forthcoming at Cambridge University Press.

Imbens, G.W. and P.R. Rosenbaum. 2005. "Robust, accurate confidence intervals with a weak instrument: quarter of birth and education." *Journal of the Royal Statistical Society Series A* 168(1):109–126.

Keele, L., C. McConnaughy and I. White. 2012. "Strengthening the Experimenters Toolbox: Statistical Estimation of Internal Validity." *American Journal of Political Science* .

Lin, Winston. 2011. "Agnostic notes on regression adjustments to experimental data: reexamining Freedman's critique." Unpublished manuscript.

McConnell, M., B. Sinclair and D.P. Green. 2010. Detecting social networks: design and analysis of multilevel experiments. In *third annual center for experimental social science and New York University experimental political science conference*.

Miguel, E. and M. Kremer. 2004. "Worms: identifying impacts on education and health in the presence of treatment externalities." *Econometrica* 72(1):159–217.

Neyman, J. 1923 [1990]. "On the application of probability theory to agricultural experiments. Essay on principles. Section 9 (1923)." *Statistical Science* 5:463–480. reprint. Transl. by Dabrowska and Speed.

Nickerson, D.W. 2008. "Is voting contagious? Evidence from two field experiments." *American Political Science Review* 102(01):49–57.

Nickerson, D.W. 2011. "Social Networks and Political Context." *Cambridge Handbook of Experimental Political Science* p. 273.

Nolen, T.L. and M. Hudgens. 2010. "Randomization-Based Inference within Principal Strata." *The University of North Carolina at Chapel Hill Department of Biostatistics Technical Report Series* p. 17.

Panagopoulos, Costas. 2006. "The Impact of Newspaper Advertising on Voter Turnout: Evidence from a Field Experiment." Paper presented at the MPSA 2006.

Rosenbaum, Paul R. 2002. *Observational Studies*. Second ed. Springer-Verlag.

Rosenbaum, Paul R. 2010. *Design of Observational Studies*. Springer.
**URL:** *http://www.springer.com/statistics/statistical+theory+and+methods/book/978-1-4419-1212-1*

Rosenbaum, P.R. 2007. "Interference Between Units in Randomized Experiments." *Journal of the American Statistical Association* 102(477):191–200.

Rosenblum, M. and M.J. Van Der Laan. 2009. "Using regression models to analyze randomized trials: Asymptotically valid hypothesis tests despite incorrectly specified models." *Biometrics* 65(3):937–945.

Rubin, D. B. 1986. "Which ifs have causal answers? comments on "Statistics and Causal Inference"." *Journal of the American Statistical Association* 81:961–962.

Rubin, Donald B. 1980. "Comment on "Randomization Analysis of Experimental Data: The Fisher Randomization Test"." *Journal of the American Statistical Association* 75(371):591–593.

Rubin, Donald B. 2005. "Causal Inference Using Potential Outcomes: Design, Modeling, Decisions." *Journal of the American Statistical Association* 100:322–331.

Samii, C. and P.M. Aronow. 2011. "On equivalencies between design-based and regression-based variance estimators for randomized experiments." *Statistics & Probability Letters* .

Sekhon, Jasjeet S. 2008. "The Neyman-Rubin Model of Causal Inference and Estimation via Matching Methods." *Oxford handbook of political methodology* pp. 271–.

Silvapulle, M.J. 1996. "A Test in the Presence of Nuisance Parameters." *Journal of the American Statistical Association* 91(436).

Sinclair, B. 2011. Design and Analysis of Experiments in Multilevel Populations. In *Cambridge Handbook of Experimental Political Science*. Cambridge University Press p. 906.

Sobel, M.E. 2006. "What Do Randomized Studies of Housing Mobility Demonstrate?" *Journal of the American Statistical Association* 101(476):1398–1407.

Tchetgen, E.J.T. and T.J. VanderWeele. 2010. "On causal inference in the presence of interference." *Statistical Methods in Medical Research* .

VanderWeele, T.J. 2008*a*. "Ignorability and stability assumptions in neighborhood effects research." *Statistics in medicine* 27(11):1934–1943.

VanderWeele, T.J. 2008*b*. "Simple relations between principal stratification and direct and indirect effects." *Statistics & Probability Letters* 78(17):2957–2962.

VanderWeele, T.J. 2009. "Marginal structural models for the estimation of direct and indirect effects." *Epidemiology* 20(1):18.

VanderWeele, T.J. 2010. "Bias formulas for sensitivity analysis for direct and indirect effects." *Epidemiology* 21(4):540.

VanderWeele, T.J. and M.A. Hernan. 2011. "Causal inference under multiple versions of treatment." *COBRA Preprint Series* p. 77.